\documentclass[10pt,journal,letterpaper,compsoc, onecolumn]{IEEEtran}

\usepackage{latexsym}
\usepackage{graphicx}
\usepackage{tabularx}
\usepackage{multirow}
\usepackage{color}
\usepackage{epsfig}
\usepackage{rotating}
\usepackage{amsmath}
\usepackage{dashrule}
\usepackage{fancyhdr}

\usepackage{hhline}
\usepackage{booktabs}
\usepackage{array}

\usepackage{hyphenat}
\usepackage[round,longnamesfirst]{natbib}

\fancypagestyle{empty}{%
  \fancyhf{}
  \fancyhead[L]{\small {\color{blue} This is a preprint of an article accepted for publication in \textit{Journal of the American Society for Information Science and Technology copyright @ 2013 (American Society for Information Science and Technology)}}}
}

 \usepackage[bookmarks=true%
,bookmarksnumbered=true%
,hypertexnames=false%
,breaklinks=true%
,colorlinks=true%
,linkcolor=blue%
,citecolor=blue%
,urlcolor=blue%
,pdfpagelabels=false
]{hyperref}

\ifCLASSOPTIONcompsoc
\else
\fi

\ifCLASSINFOpdf
\else
\fi

\ifCLASSOPTIONcompsoc
\usepackage[tight,normalsize,sf,SF]{subfigure}
\else
\usepackage[tight,footnotesize]{subfigure}
\fi

\begin{document}
\pagenumbering{gobble}

\title{Real-Time Classification of Twitter Trends}

\author{
  Arkaitz Zubiaga$^1$, Damiano Spina$^2$$^*$, Raquel Mart\'{i}nez$^2$, V\'{i}ctor 
Fresno$^2$ \\
  $^1$ Dublin Institute of Technology \\  DIT Focas Institute, Camden Row \\ Dublin 8, Ireland \\
  $^2$ NLP \& IR Group at UNED \\ C/ Juan del Rosal, 16 \\ 28040 Madrid, Spain \\
  $^*$ Corresponding author. \\
  {\tt arkaitz@zubiaga.org, \{damiano,raquel,vfresno\}@lsi.uned.es} \\
  
}

\IEEEcompsoctitleabstractindextext{%
\begin{abstract}
Social media users give rise to social trends as they share about common interests, which can be triggered by different reasons. In this work, we explore the types of triggers that spark trends on Twitter, introducing a typology with following four types: \emph{news}, \emph{ongoing events}, \emph{memes}, and \emph{commemoratives}. While previous research has analyzed trending topics in a long term, we look at the earliest tweets that produce a trend, with the aim of categorizing trends early on. This would allow to provide a filtered subset of trends to end users. We analyze and experiment with a set of straightforward language-independent features based on the social spread of trends to categorize them into the introduced typology. Our method provides an efficient way to accurately categorize trending topics without need of external data, enabling news organizations to discover breaking news in real-time, or to quickly identify viral memes that might enrich marketing decisions, among others. The analysis of social features also reveals patterns associated with each type of trend, such as tweets about ongoing events being shorter as many were likely sent from mobile devices, or memes having more retweets originating from a few trend-setters.
\end{abstract}

}

\maketitle

\IEEEdisplaynotcompsoctitleabstractindextext

\IEEEpeerreviewmaketitle

\thispagestyle{empty}

\section*{Introduction}
\label{introduction}

While social media have become mainstream, as shown by the evergrowing number of users, Twitter\footnote{\url{http://twitter.com/}} stands out as the quintessential platform to openly access real-time updates on breaking news and ongoing events. With over 500 million users, Twitter sees a daily stream of more than 400 million short messages known as tweets\footnote{\href{http://expandedramblings.com/index.php/march-2013-by-the-numbers-a-few-amazing-twitter-stats/}{http://expandedramblings.com/index.php/march-2013-by-the- numbers-a-few-amazing-twitter-stats/}}. These tweets include from one-to-one conversations and chatter, to updates of wider interest about current affairs, encompassing all kinds of information.

One of the appealing phenomena of the microblogging service is the fact that certain occurrences of wide interest for a community of users produce a sudden increase of mentions in real-time as they unfold. Interestingly, users live-tweet about sporting events as they watch them on TV, discuss breaking news as they learn about them, or commemorate certain events on a memorial day, among others. These result in a spiky activity associated with the occurrence in question, which produces what is known as a social trend. While these social trends can reveal what is going on at the very moment, and discover certain events and breaking news early on, a list of social trends includes just the set of terms that are being mentioned outstandingly at the very moment, but no context is provided to learn more about what triggered the social trend as well as the kind of event behind each of the trends. Little attention has been paid so far to researching social trends so as to mine additional characteristics from them and 
understand why they were produced. Finding out the trigger that produces a social trend can help not only 
inform users about social trends of their interest, but also feed third parties with different interests: for 
instance, news media can be interested in early discovery of social trends associated with breaking news 
\citep{zubiaga2013curating}, governments could be interested in tracking issues concerning certain events 
for security issues, and marketing professionals might be interested in quickly identifying viral memes to 
react accurately. In this context, previous research on the analysis of social trends has focused on 
long-term analysis of social conversations, but little is known about the origin of social trends with the 
aim of understanding and categorizing them by type to enable real-time filtering.

In this work, we set out to develop and experiment with an automatic classifier to identify the type of 
trigger that produces a social trend. This classifier is intended to quickly categorize trends as they emerge, 
allowing to inform end users about social trends in real-time in a filtered way. We introduce a typology of 
trends we came up with after carefully tracking and analyzing social trends for a long time, which encompasses 
the different kinds of triggers and enables to provide different communities with the trends of their 
interest. Based on our hypothesis that different types of trends will present different patterns in terms of 
generation and social diffusion, we define an automatic classification method that complies with the following 
assets to be used efficiently in real-time: (i) it requires a small and language-independent set of features 
that can be straightforwardly 
obtained, (ii) it does not make use of external data, (iii) it performs accurately outperforming a baseline 
approach relying on the content of tweets, and (iv) it requires low computational cost. We complement the 
study with an analysis of the distribution of the features we defined, bringing to light behavioral patterns 
for each kind of social trend, which explains how information 
is spread in each case. 

The remainder of this paper is organized as follows. Next, we provide background 
on the use of Twitter, the syntax utilized by its users, and trends. Then, we 
introduce a typology to organize trends by type of trigger, and detail the dataset we generated to carry out 
the analysis and experiments. We define and analyze a set of features for characterizing trending topics and 
describe the trend classification experiments and results. Finally, we discuss the contributions of the 
study, contextualizing it with the related work and wrapping up with the conclusions.

\section*{Background}
\label{background}

In this section, we provide background on Twitter relevant to this work, describing the syntax utilized by its 
users, and the way they interact with each other and spread information. Then, we get into detail of how 
Twitter presents and deals with trends (defined as \textit{trending topics} by the microblogging system), 
which we utilize as the input to our trend classifier.

\subsection*{Twitter}
\label{twitter}

Twitter has become a huge social media service where millions of users contribute on a daily basis. Two 
features have been fundamental in its success: (1) the shortness of tweets, which cannot exceed 140 
characters, facilitates creation and sharing of messages in a few seconds, and (2) the easiness of spreading 
those messages to a large number of users in very little time. Throughout the time, the community of users on 
Twitter has established a syntax for interaction with one another, which has become the standard syntax later 
officially adopted by its developers. Most major Twitter clients have implemented this standard syntax as 
well. The standards in the interaction syntax include:

\begin{itemize}
 \item \textbf{User mentions:} when a user mentions another user in their tweet, an at-sign is placed before 
the corresponding username, e.g., \texttt{You should all follow @username, she is always abreast of breaking 
news and interesting stuff}.

 \item \textbf{Replies:} when a user wants to direct to another user, or reply to an earlier tweet, they place 
the \texttt{@username} mention at the beginning of the tweet, e.g., \texttt{@username I agree with you}.

 \item \textbf{Retweets:} a retweet is considered a re-share of a tweet posted by another user, i.e., a 
retweet means the user considers that the message in the tweet might be of interest to others. When a user 
retweets, the new tweet copies the original one in it. Furthermore, the retweet attaches an \texttt{RT} and 
the \texttt{@username} of the user who posted the original tweet at the beginning of the retweet. For 
instance: if the user \texttt{@username} posted the tweet \texttt{Text of the original tweet}, a retweet on 
that tweet would look this way: \texttt{RT @username: Text of the original tweet}. Moreover, retweets can 
further be retweeted by others, what creates a retweet of level 2, e.g., \texttt{RT @username2: RT @username: 
Text of the original tweet}. Similarly, retweets can go deeper into 3rd level, 4th, and so forth.

 \item \textbf{Hashtags:} similar to tags on social tagging systems or other social networking systems, 
hashtags included in a tweet tend to group tweets in conversations or represent the main terms of the tweet, 
usually referred to topics or common interests of a community. A hashtag is differentiated from the rest of 
the terms in the tweet in that it has a leading hash, e.g., \texttt{\#hashtag}.

\end{itemize}

\subsection*{Trending Topics}
\label{trending-topics}

One of the main features on the homepage of Twitter shows a list of top terms so-called trending topics at all 
times. These terms reflect the topics that are being discussed most at the very moment on the site's 
fast-flowing stream of tweets. In order to avoid topics that are popular regularly (e.g., \textit{good 
morning} or \textit{good night} on certain times of the day), Twitter focuses on topics that are being 
discussed much more than usual, i.e., topics that recently suffered an increase of use, so that it trended for 
some reason. Trending topics have attracted big interest not only among the users themselves but also among 
other information consumers such as journalists, real-time application developers, and social media 
researchers. Being able to know the top conversations being discussed at a given time helps keep updated about 
current affairs, and discover the main concerns of the community. Twitter defines trending topics as 
\emph{``topics that are immediately popular, rather than topics that have 
been popular for a while or on a daily 
basis''}\footnote{\url{http://support.twitter.com/articles/101125-about-trending-topics}}. However, no 
further evidence is known about the algorithm that extracts trending topics. It is assumed that the list is 
made up by terms that appear more frequently in the most recent stream of tweets than the usual expected 
\citep{asur2011trends}.

A trending topic is made up by the topic itself --i.e., the term or phrase that became a trend--, and a stream 
of tweets containing that topic. Table \ref{tab:tt-tweets} shows an example of a trending topic and some 
underlying tweets. In this example, @u1 posted an early tweet reporting a news, which was retweeted by @u2, 
and by @u4 afterward; @u3 replied to @u1 by asking further information about the news, and @u5 posted a new 
tweet that points at a link to the news.

\begin{table}[htb]
 \begin{center}
  \begin{tabular}{ | l p{6cm} | }
   \hline
   \multicolumn{2}{ | c | }{\textbf{Trending Topic:} Interpol} \\
   \hline
   \textbf{User} & \textbf{Content of the tweet} \\
   \hline
   @u1 & Interpol issues arrests warrants for Gaddafi \& 15 senior Libyan officials. \#Libya \\
   \hline
   @u2 & RT @u1: Interpol issues arrests warrants for Gaddafi \& 15 senior Libyan officials. \#Libya \\
   \hline
   @u3 & @u1 - so Interpol cannot act until he \& family leave Libya, is that right? Assuming he is toppled? 
\\
   \hline
   @u4 & RT @u2: RT @u1: Interpol issues arrests warrants for Gaddafi \& 15 senior Libyan officials. \#Libya 
\\
   \hline
   @u5 & Interpol has issued international alert for Muammar Gaddafi \& 15 other family members \& close 
associates | Telegraph http://bit.ly/h9GwYI \\
   \hline
  \end{tabular}
 \end{center}
 \caption{Example of a trending topic, and some tweets associated.}
 \label{tab:tt-tweets}
\end{table}

In this work, we rely on trending topics as provided by Twitter through its API, therefore controlling for the 
detection of trends, so we can focus on the classification of social trends once they have already been 
detected.

\section*{Typology of Trending Topics}
\label{trend-types}

Next, we introduce the typology we use in this work to organize trending topics. After describing the 
typology, we detail the process of generation of a dataset, made up by trending topics and organized according 
to the typology.

\subsection*{Definition of a Typology of Trending Topics}
\label{identification-of-types}

Not all the trending topics emerge equally. Different happenings, either in the society, on a TV show, or on 
the Internet, can motivate users to share and discuss on a topic, producing a sudden increase of associated 
sharings and subsequently the creation of a trending topic. After coming up with this hypothesis, we started 
tracking Twitter's trending topics while we were users of the microblogging service. This started as an 
informal process, where we checked the trending topics occasionally to see what was going on at the moment. 
When we noticed that very different kinds of events were interspersed in trending topics, we proceeded with a 
rather formal process of creating a list of potential categories. After discussing and defining a typology 
agreed by the four researchers, we continued checking trending topics for another month to make sure that the 
typology covered them all. This enabled us to establish the typology of trending topics that encompasses the 
types of conversations that arouse the interest 
of large communities of users in short time periods by producing trending topics. The typology we defined is 
composed by the following four types of triggers:

\begin{itemize}
 \item \textbf{News:} breaking news tend to make it to Twitter early on, having even shown that on many 
occasions news break on Twitter before any news outlets report it \citep{kwak2010what}. In the course of 
trend monitoring, we saw that news do indeed represent a subset of trending topics in the early stage of a 
breaking news. We can define that a trending topic can be categorized as \textit{news} when it is produced by 
a newsworthy event that major news outlets either had reported it by the time the trend popped up or will 
report it soon after it broke on Twitter.

 \item \textbf{Ongoing events:} another type of trending topic we identified was produced by a community of 
users tweeting about an ongoing event as it unfolds. The practice of live-tweeting an event as it is taking 
place has become fundamental as Twitter has gained importance as a real-time information sharing media. Here 
an event can be from a soccer game to a keynote presentation by Apple, a music festival, or a conference.

 \item \textbf{Memes:} we also identified that a portion of trending topics were triggered by viral ideas 
initiated by either an individual or an organization, who were usually popular enough to be able to spread 
something widely. We have labeled these type of trending topics as \textit{memes}, which we can define as the 
event that, without being apparently newsworthy or a mainstream event that a large audience is following, 
makes it to a large community of users for being funny or attractive to them. It can be from a funny message 
by a teen heartthrob such as Justin Bieber, to a protest leader's request to spread a message in support of a 
plea or complaint.

 \item \textbf{Commemoratives:} the last type of trending topic we identified, which was probably the least 
frequent, was that produced by the commemoration of certain person or event that is being remembered in a 
given day. We define a trending topic as triggered by a \textit{commemorative} when users are congratulating a 
celebrity for their birthday, celebrate the anniversary of a certain event or person, or it is a memorial day.
\end{itemize}

Having tracked trending topics for a long period of time while we kept tweaking this typology, we believe 
that 
it can be considered as a cross-grained categorization that encompasses a vast majority of trending topics as 
of today on Twitter. While we believe that the presented typology includes the majority of triggers producing 
trends on Twitter, we did also realize that certain trends could fall into more than one of the categories 
described above. Even though the categorization of trends could be multi-label, assigning more than one 
category to each trend when applies, in this work we set out to choose the category in which a trend fits 
best. For instance, we consider that Liverpool supporters cheering for their soccer team as soon as 
they scored 
against Manchester United can be categorized as an \textit{ongoing event}, and we consider as a \textit{news} 
the announcement of the Defense Minister's retreat in Germany, given that it is a newsworthy event that will 
be reported by news media right away.

Table \ref{tab:typology-examples} depicts this typology with an example of a trending topic for each of the 
four types.

\begin{table*}[htb]
\small{
 \begin{center}
  \begin{tabular}{ l l p{5cm} p{6cm} }
   \toprule
   \textbf{Type} & \textbf{Trending topic} & \textbf{Description} & \textbf{Sample tweet} \\
   \midrule[.5pt]
   News & \texttt{Islamabad} & Minority Minister of Pakistan was murdered in Islamabad & \texttt{RT 
@BreakingNews: Pakistani Minister for Minorities Shahbaz Bhatti killed in a gunattack in Islamabad - AP} \\
   Ongoing event & \texttt{Anfield} & A soccer game was being played at Anfield, the home stadium of Liverpool 
FC, an English soccer team & \texttt{Liverpool vs manchester united @ Anfield :D} \\
   Meme & \texttt{\#5bestsoundtracks} & People were tweeting their list of the 5 best soundtracks, as per 
request of the NME music magazine & \texttt{@NMEmagazine 1. Kill Bill, 2. Trainspotting, 3. Across the 
Universe, 4. Juno, 5. This is England \#5bestsoundtracks} \\
   Commemorative & \texttt{\#worldbookday} & People were commemorating that it was the World Book Day & 
\texttt{happy \#worldbookday everyone! what are you reading?} \\
   \bottomrule
  \end{tabular}
 \end{center}
}
 \caption{Examples of trending topics for the presented typology.}
 \label{tab:typology-examples}
\end{table*}

Having trending topics categorized in this typology enables not only to deliver them to end users interested 
in specific categories, but also to feed systems that benefit from the input of trending topics with suitable 
categories.

\subsection*{Dataset}
\label{dataset}

Twitter employs an algorithm that ultimately displays the top 10 list of terms that are being discussed most 
at the moment. This list of top 10 terms is shown in the trending topics section on the homepage of the site, 
which is also available through its API. The list of trending topics obtained in real-time through the API can 
be complemented with another method in the API that allows to search for tweets containing a given query term, 
where the query term can be each of these trending topics. The search API applies to all the public tweets 
posted by any user on the site, and it returns up to 1,500 of the most recent tweets. Using the top trends and 
search API methods, we monitored the trending topics shown on the site from March 1st to 7th, 2011. We had two 
processes running at the same time in order to gather both trending topics and the underlying tweets. The 
first process worked on the top trends API, by monitoring the trending topics appeared on the site. It 
requested the list of top 10 trending topics 
every 
30 seconds. Thus, the process guarantees the detection of a trending topic almost as soon as it appears on the 
site, with a delay 
of 30 seconds in the worst case scenario. The process checked the gathered list for new trends that were not 
seen in previous requests. As soon as a new topic appeared in the list of trends, the second process queried 
the search API for the latest tweets containing the topic as a query term. It gathered thus a list of recent 
tweets that included the trending topic. Among the gathered information, we saved all the trending topics, and 
the text, timestamp, user, and language for each of the underlying tweets.

Following the above gathering process, we collected a total of 1,036 unique trending topics. These trends 
include a total of 567,452 tweets from 348,757 different users. Accordingly, each of the 1,036 trending topics 
in the dataset contains an average of about 548 associated tweets. All these tweets are written in 28 
different languages, with a majority of 295,082 tweets written in English, 76,628 in Spanish, 67,673 in 
Portuguese, 31,685 in Dutch, and 22,863 in Indonesian. The stream of a trending topic can also include tweets 
written in different languages. In fact, there is not a trending topic with just one language in the dataset; 
403 topics have tweets written in 28 languages, 232 have tweets in 27 languages, and 320 have tweets in 26 
languages. There are just 2 trending topics with tweets in only 8 languages, and the rest include tweets in at 
least 12 languages. This is an evidence of the spread power of conversations on Twitter, insofar as many 
languages (thus assumedly many countries) contribute to 
share it. Taking into account the predominant language of tweets contained in each trending topic, 668 trends 
have a majority of tweets in English, 131 in Portuguese, 118 in Spanish, 40 in Dutch, 19 in Indonesian, 18 in 
German, and the remainder 42 belong to 12 different languages.

In order to perform the classification experiments, and to analyze the characteristics of the gathered 
trending topics, all these topics were manually annotated according to its type within the typology we 
defined: news, ongoing events, memes, or commemoratives. Four people, all of them being everyday Twitter 
users, performed the annotation process, where they had to assign each trending topic a single category from 
the aforementioned four types. The annotation process was divided randomly in such a way that each trending 
topic was annotated by three of the annotators. The annotators carefully read through the tweets in a 
trending 
topic, so they could understand what was going on, assigning the category that better suited. When tweets 
were 
written in a language that the annotator in question could not understand, Google's translation 
system\footnote{\url{http://translate.google.com/}} was used to facilitate the task. The annotation process 
produced the following distribution of categories: 616 
ongoing 
events 
(59.5\%), 251 memes (24.2\%), 142 news (13.7\%), and 27 commemoratives (2.6\%). In order to 
measure the inter-annotator agreement we computed the Fleiss' Kappa coefficient~\citep{fleiss1971measuring}, 
obtaining a value of 0.597. According to~\citet{landis1977agreement}, kappa values between 0.4 and 0.6 
may be taken to represent 'moderate' agreement beyond chance~\footnote{Note that the range of 
possible values for Fleiss' Kappa coefficient includes values below zero. In this case, a 0.597 Kappa is the 
equivalent to 90.5\%
agreement rate among annotators.}.

The trending topics and its annotations used in this work are publicly available for 
further research\footnote{\url{http://nlp.uned.es/\~damiano/datasets/TT-classification.html}}.

\section*{Characterization of Trending Topics}

\label{characterization-trending-topics}

Next, we present the classification experiments as well as the feature analysis, with which we evaluate and 
quantify the validity of the introduced typology. We first analyze and discuss the features, and then present 
classification experiments based on those features.

\subsection*{Features of Trending Topics}
\label{tt-features}

The main motivation of this work is our hypothesis that depending on the type of trigger that produces a 
trend, the social patterns observed on users' behavior in terms of information diffusion will vary. Having 
this hypothesis as a motivation and starting point, we define a set of 15 social features to help capture 
this social behavior. We analyze the extent to which these features identify behavioral patterns so as to 
accurately categorize trends in real-time. Later, we dig into these social features to extract insight about 
what they mean in each type of trend.

As an approach to discovering the type of trigger behind a trending topic, we propose 15 social features that 
consider the way it spreads. Furthermore, since we want to categorize a trending topic as soon as it appears 
trending on the site, the features must be straightforward, easy to get, and cheap to compute while the 
system performs accurately. Note that the number of tweets in a topic that just trended is relatively small 
(average of 548 tweets), thus computing the features is fast and computationally cheap. This would ensure the 
immediacy of the computation, and the ability to predict the type of a trending topic on the fly, as soon as 
it appears on the Twitter's list. Moreover, these features are independent of the language used in tweets, and 
do not depend on the vocabulary utilized by users.

We believe that each type of trending topic spreads in a different way, and that social features have much to 
do with the type. Our conjecture is that trending topics brought about by different types of trigger will be 
socially shared in a different way. For instance, one could expect that tweets about news tend to be 
accompanied by links more than ongoing events, or that memes could be shared and spread in a different way as 
compared to other types.

The 15 social features we propose for characterizing trending topics are shown in Table \ref{tab:features}. 
The table shows both the name and an explanatory definition for each feature. The features are organized in 
two groups, according to the type of calculation used.

On one hand, we use average number of occurrences of features in the tweets corresponding to a trend. Each 
average computed as the arithmetic mean is the result of dividing the number of occurrences of the 
corresponding feature in the whole trend by the total number of tweets gathered for the trending topic (see 
Equation \ref{eq:average}).

\begin{equation}
 AM(f)_t = \frac{1}{|T|} \sum_{i=1}^{|T|} f_i
 \label{eq:average}
\end{equation}

where $AM(f)_t$ is the arithmetic mean of feature $f$ for the trending topic $t$, $|T|$ is the number of 
tweets in the trending topic, and $f_i$ is the value of feature $f$ for the tweet $i$.

We propose 10 different features that rely on average values: \emph{Retweets (depth)}, \emph{Retweets 
(ratio)}, \emph{Hashtags}, \emph{Length}, \emph{Exclamations}, \emph{Questions}, \emph{Links}, \emph{Topic 
repetition}, \emph{Replies}, and \emph{Spread velocity}. All these features can be computed by following 
Equation \ref{eq:average}, except the \emph{Spread velocity}, which follows a slightly different equation to 
compute the average number of tweets posted per second (see Equation \ref{eq:average-sv}).

\begin{equation}
 AM(sv)_t = \frac{|T|}{\Delta t}
 \label{eq:average-sv}
\end{equation}

where $\Delta t=$ is total number of seconds from the first to the last tweet in the trending topic.

On the other hand, we compute the diversity of feature values all across the tweets that belong to a trending 
topic. The diversity calculates the variation of the feature throughout the trending topic. The higher is the 
diversity value, the more different is the feature from tweet to tweet within a trending topic. To compute 
the diversity, we use Shannon's diversity index (see Equation \ref{eq:shannon-index}). In other words, 
Shannon's index is the information entropy of the distribution of values for a feature, i.e., each different 
value is considered a symbol, and the population of each value is used as the probability.

\begin{equation}
 H'(f)_t = - \sum_{j = 1}^{S} (p_{jt} \ln p_{jt});\mbox{   }p_{jt} = \frac{n_{jt}}{N}
 \label{eq:shannon-index}
\end{equation}

where $H'(f)_t$ is the Shannon's index of feature $f$ for the trending topic $t$, $n_{jt}$ is the population 
of the value $j$, $S$ is the number of different values, $N$ is the total population, and $p_{jt}$ is the 
observed probability of the value $j$.

We propose 5 features that rely on diversity: \emph{User diversity}, \emph{Retweeted user diversity}, 
\emph{Hashtag diversity}, \emph{Language diversity}, and \emph{Vocabulary diversity}.

\begin{table*}[htb]
 \begin{center}
  \begin{tabular}{ l p{12cm} }
    \toprule[1pt]
    \multicolumn{2}{c}{\textbf{Averages / Arithmetic Means}} \\
    \midrule[.5pt]
    \textbf{Feature} & \textbf{Definition} \\
    \midrule[.5pt]
    \vspace{5pt}
    \textbf{Retweets (depth)} & Average number of retweet levels in tweets. The retweet level of a tweet is 
the number of retweeting users before it reached the current state. For instance, the retweet of an original 
tweet corresponds to level 1, the retweet of a retweet to level 2, and so on. This feature adds up all these 
levels, and divides it by the total number of tweets. \\
    \vspace{5pt}
    \textbf{Retweets (ratio)} & Ratio of tweets that contain a retweet. It is the result of dividing the 
number of tweets that are retweets (regardless of the level) by the total number of tweets. \\
    \vspace{5pt}
    \textbf{Hashtags} & Average number of hashtags in tweets. \\
    \vspace{5pt}
    \textbf{Length} & Average length of tweets, i.e., the average number of characters used. \\
    \vspace{5pt}
    \textbf{Exclamations} & Number of tweets with exclamation signs. Tweets with at least an exclamation are 
given a value of 1, whereas tweets without exclamations are given a value of 0. This feature computes the 
average of those values. \\
    \vspace{5pt}
    \textbf{Questions} & Number of question signs in tweets. It is the same as with exclamations, but with 
question signs. \\
    \vspace{5pt}
    \textbf{Links} & Average number of links in tweets \\
    \vspace{5pt}
    \textbf{Topic repetition} & Average number of uses of the trending topic in tweets. The term 
corresponding to the trending topic may appear more than once in a tweet, especially when users repeat it to 
make it stronger and trend. \\
    \vspace{5pt}
    \textbf{Replies} & Average number of tweets that are replies to others. A tweet is given a value of 1 
when it is a reply to another user, or a value of 0 otherwise. \\
    \vspace{5pt}
    \textbf{Spread velocity} & Average number of tweets per second in the trend. It is the result of dividing 
the total number of tweets by the period of time between the last and first tweets in the trend. \\
    \midrule[1pt]
    \multicolumn{2}{c}{\textbf{Diversity values}} \\
    \midrule[.5pt]
    \textbf{Feature} & \textbf{Definition} \\
    \midrule[.5pt]
    \vspace{5pt}
    \textbf{User diversity} & Shannon's diversity index of users who posted tweets. When a few users are 
repeatedly tweeting on the topic, the diversity decreases. \\
    \vspace{5pt}
    \textbf{Retweeted user diversity} & Shannon's diversity index of users who were retweeted in the trend 
(not the user who retweets). The diversity value is higher when different users get retweeted, and lower when 
retweets come from a few users. \\
    \vspace{5pt}
    \textbf{Hashtag diversity} & Shannon's diversity index of hashtags included in the trend. When the same 
hashtag is used throughout a trend, the diversity decreases. \\
    \vspace{5pt}
    \textbf{Language diversity} & Shannon's diversity index of languages used in the trend. The diversity is 
lower when most of the tweets are written in the same language, and higher when several languages have some 
tweets. \\
    \vspace{5pt}
    \textbf{Vocabulary diversity} & Shannon's diversity index of terms contained in the trend. The diversity 
decreases when tweets share a similar vocabulary, whereas it increases the vocabulary varies among tweets. \\
    \bottomrule[1pt]
  \end{tabular}
 \end{center}
 \caption{List of Twitter features and definitions.}
 \label{tab:features}
\end{table*}

\subsection*{Understanding the Features}
\label{trend-analysis}

Next, we analyze and discuss the studied features, not only for the Twitter
features, but also for the textual content of tweets. To do so, we explore the distributions of the Twitter
features for each type of trending topic, and dig into the main terms that characterize each of the types.
This helps understand whether and why the features we used discriminate among types of trending topics.

\subsubsection*{Analysis of Twitter Features}
\label{twitter-feature-analysis}

The use of Twitter features produced an accurate performance for classifying trending topics according to the
proposed typology. Going further, it is of utmost importance to analyze how the values of these features
distribute for each type of trending topic, and the extent to which they can help discriminate types. We rely
on box/whisker plots to show these distributions of values, and differences or similarities between types.
Each box/whisker plot shows the distribution of values of a feature for the 4 types of trending topics. The
line in the middle of the box is the median (second quartile), the bottom of the box is the first quartile,
the top of the box is the third quartile, and the top bar is the maximum value observed where the bottom bar 
is the
lowest. This kind of plot helps visualize the range of values for each feature, as well as see where most of
the values congregate. It enables to perform a qualitative analysis of the features.

\begin{figure*}[htb]
 \centering

 \subfigure[Retweets (depth)]{
  \includegraphics[width=90px]{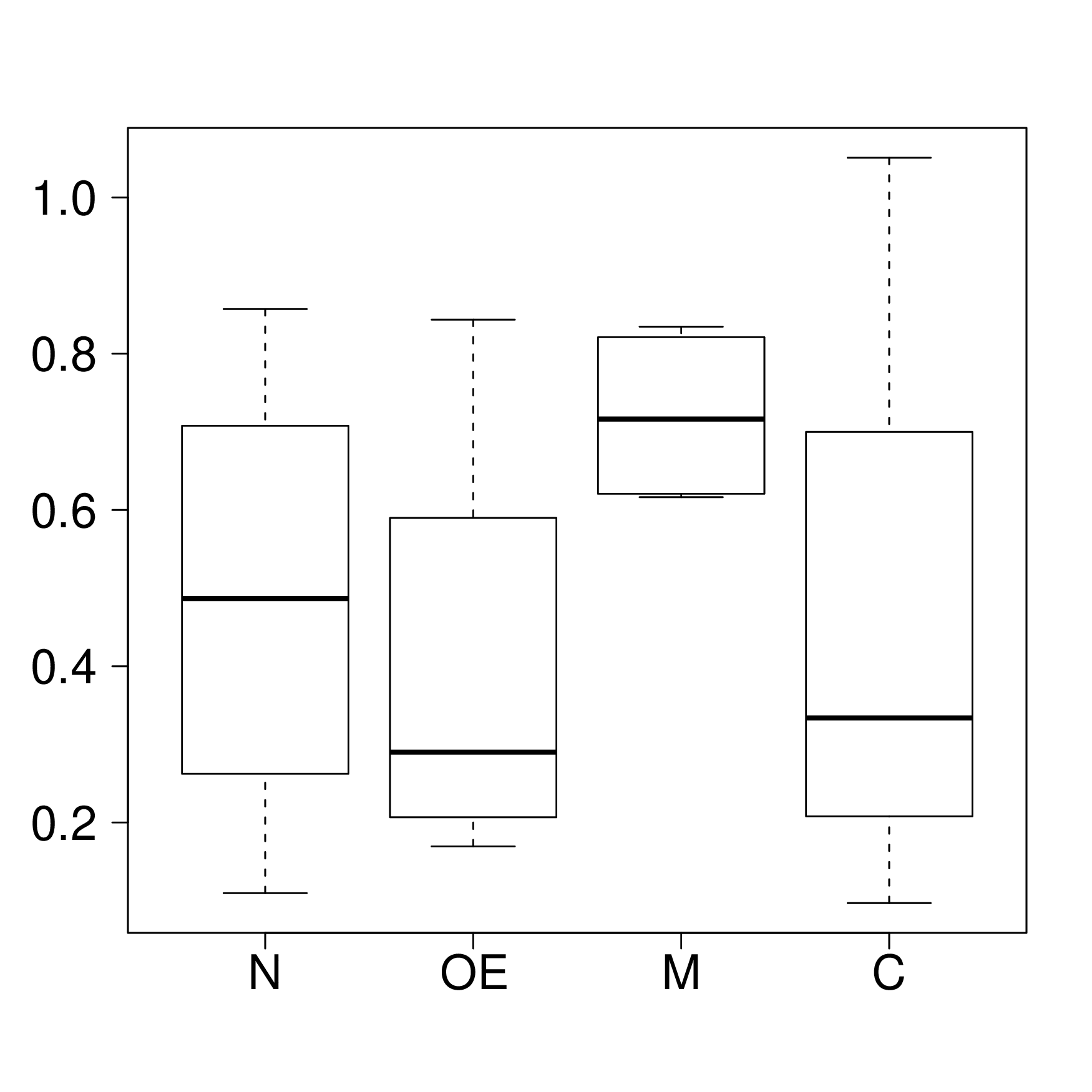}
  \label{fig:boxplot-retweets}
 }
 \subfigure[Retweets (ratio)]{
  \includegraphics[width=90px]{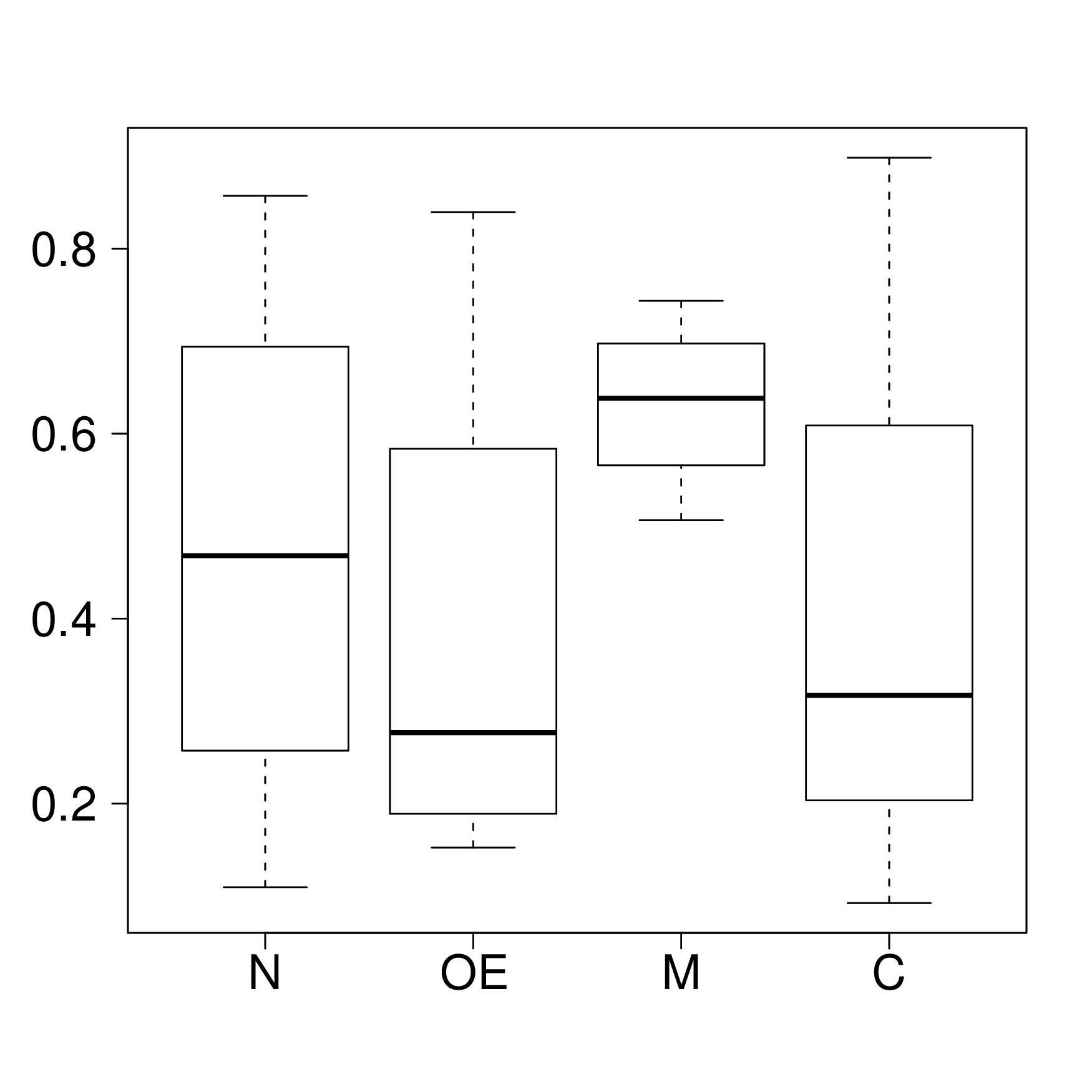}
  \label{fig:boxplot-retweets-ratio}
 }
 \subfigure[Hashtags]{
  \includegraphics[width=90px]{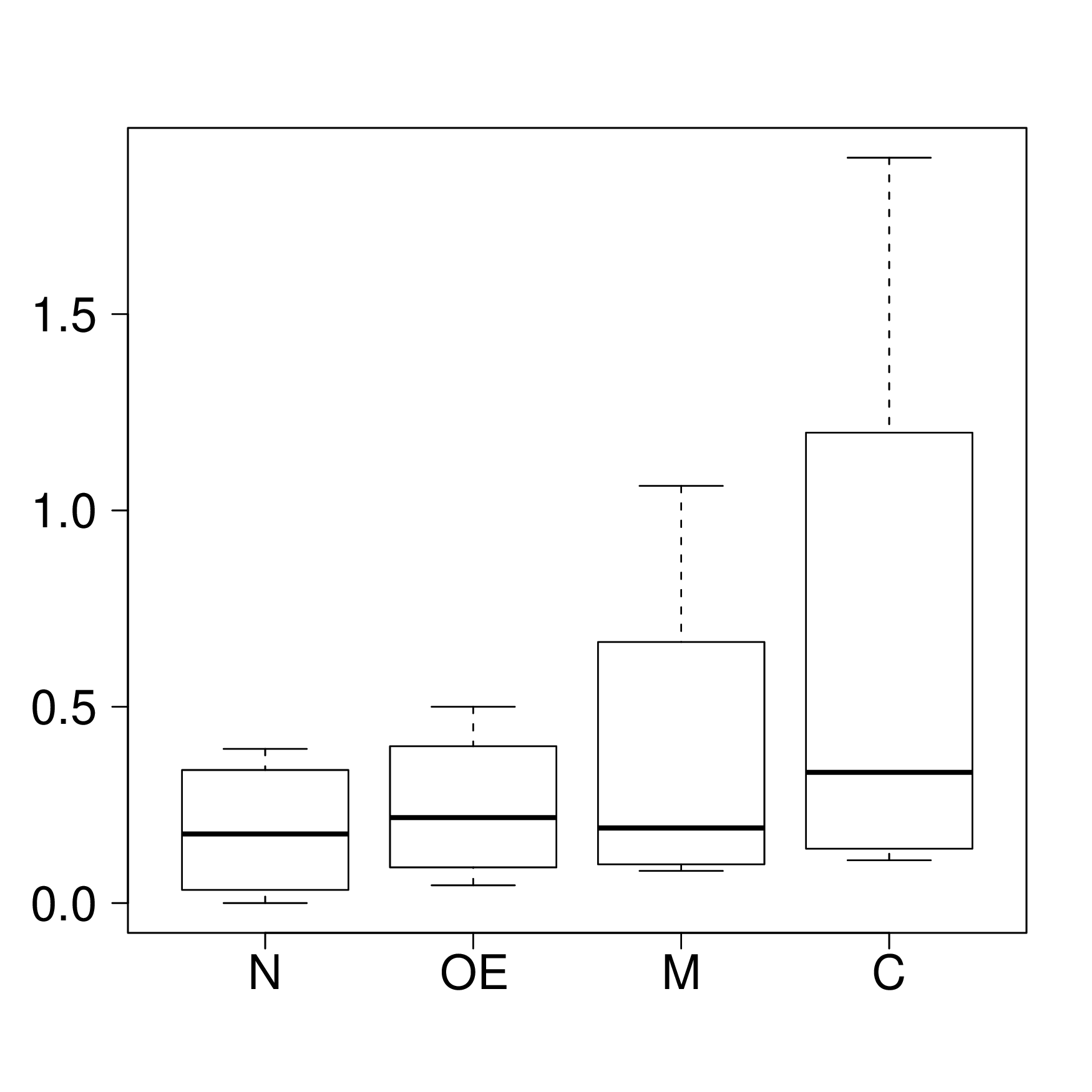}
  \label{fig:boxplot-hashtags}
 }
 \subfigure[Length]{
  \includegraphics[width=90px]{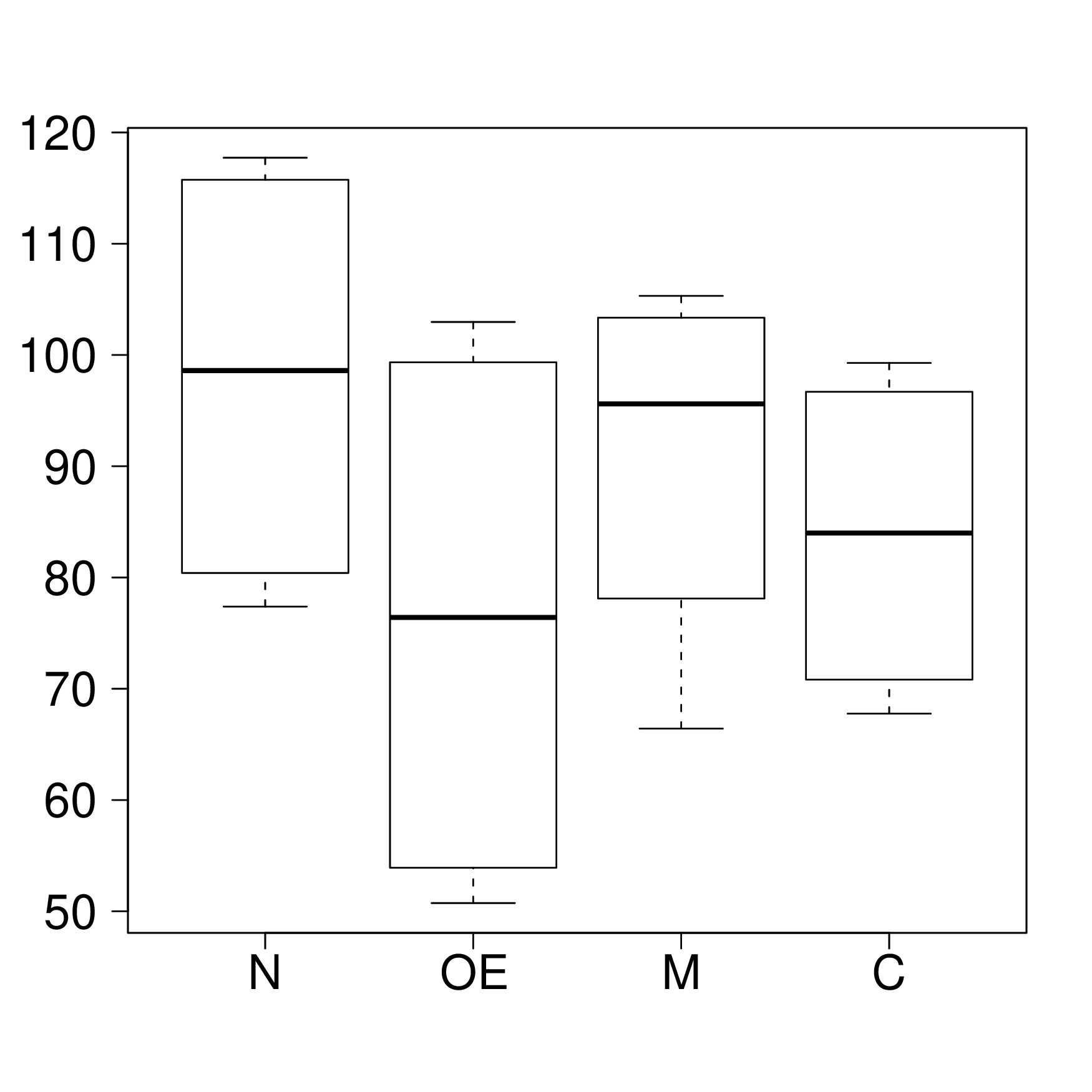}
  \label{fig:boxplot-length}
 }
 \subfigure[Exclamations]{
  \includegraphics[width=90px]{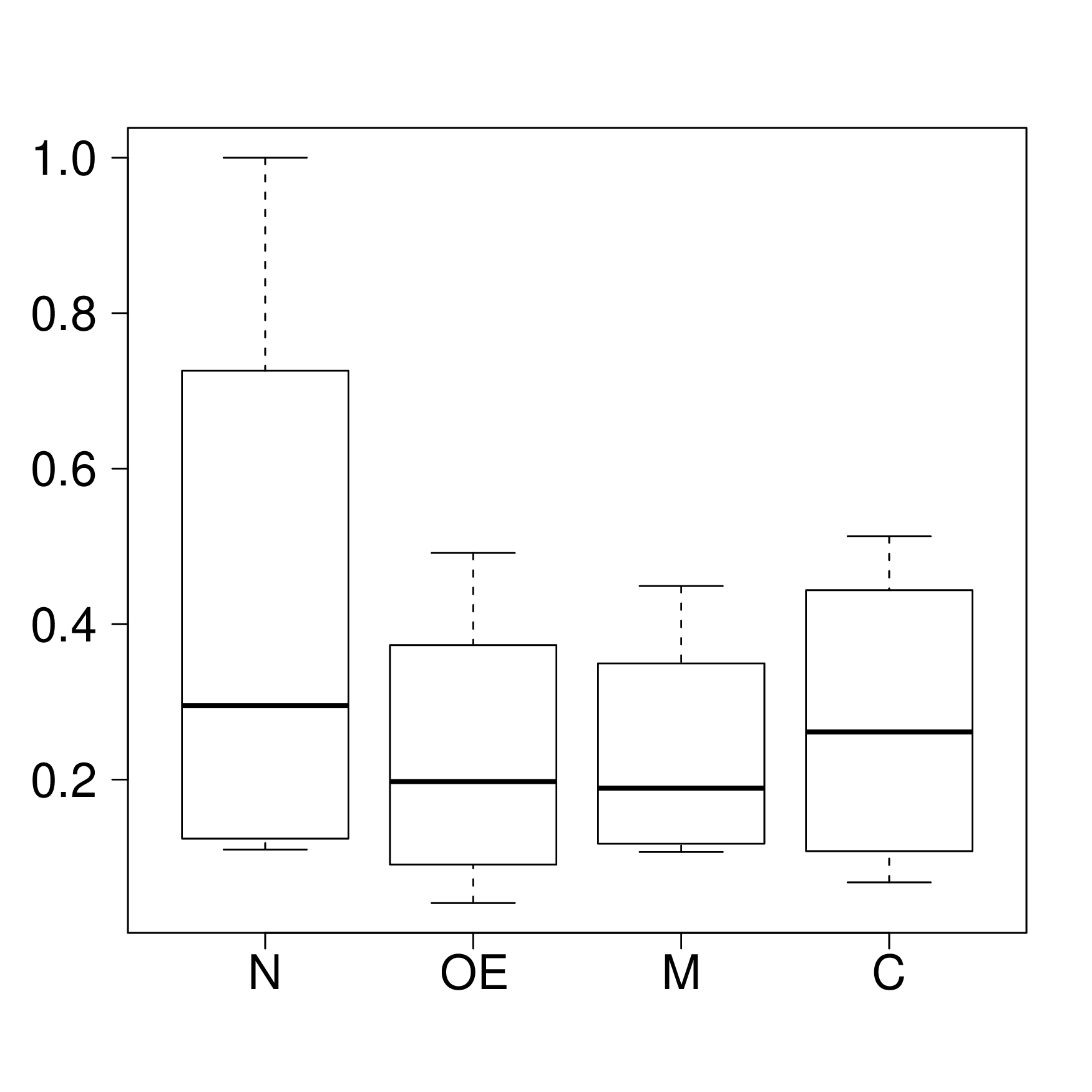}
  \label{fig:boxplot-exclamations}
 }
 \subfigure[Questions]{
  \includegraphics[width=90px]{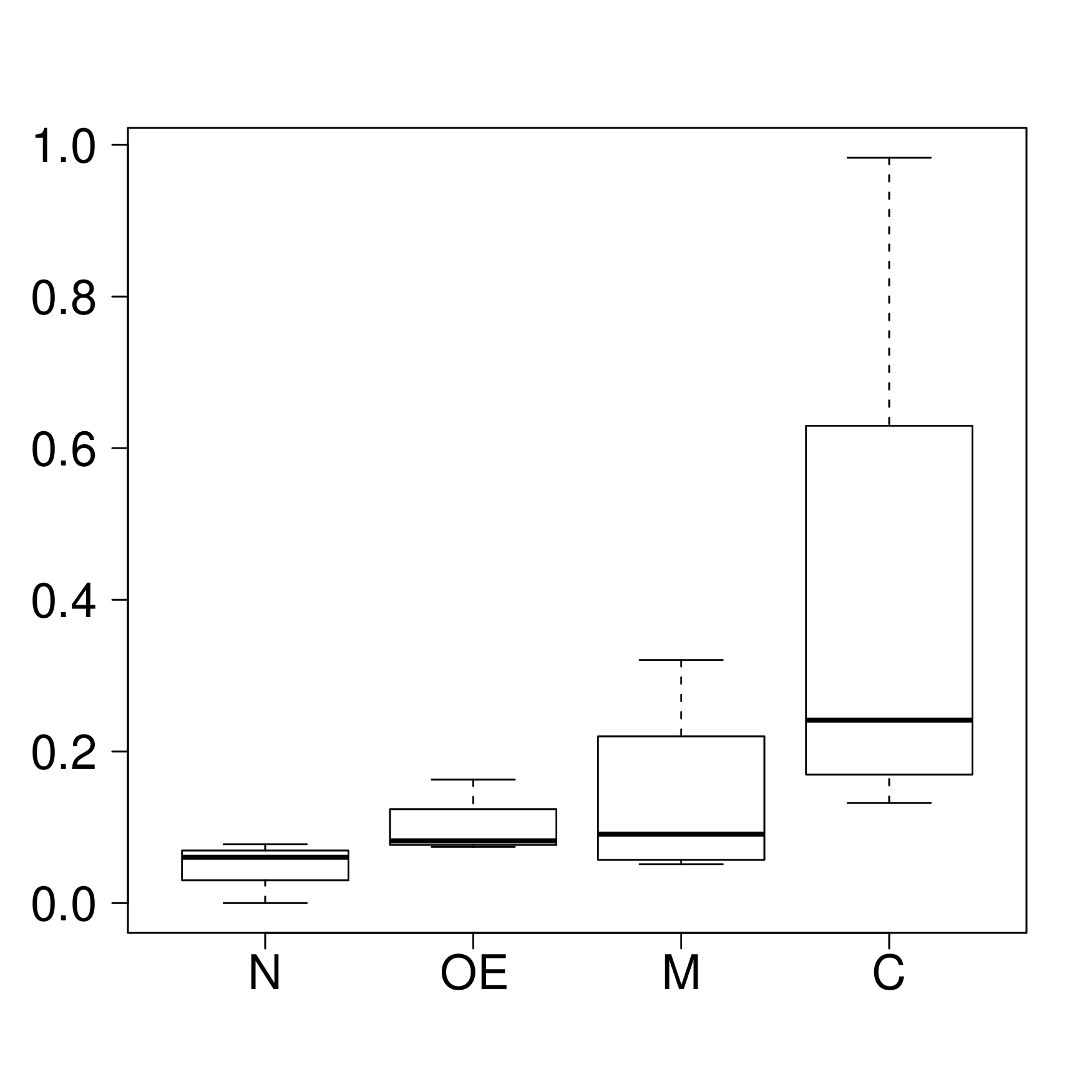}
  \label{fig:boxplot-questions}
 }
 \subfigure[Links]{
  \includegraphics[width=90px]{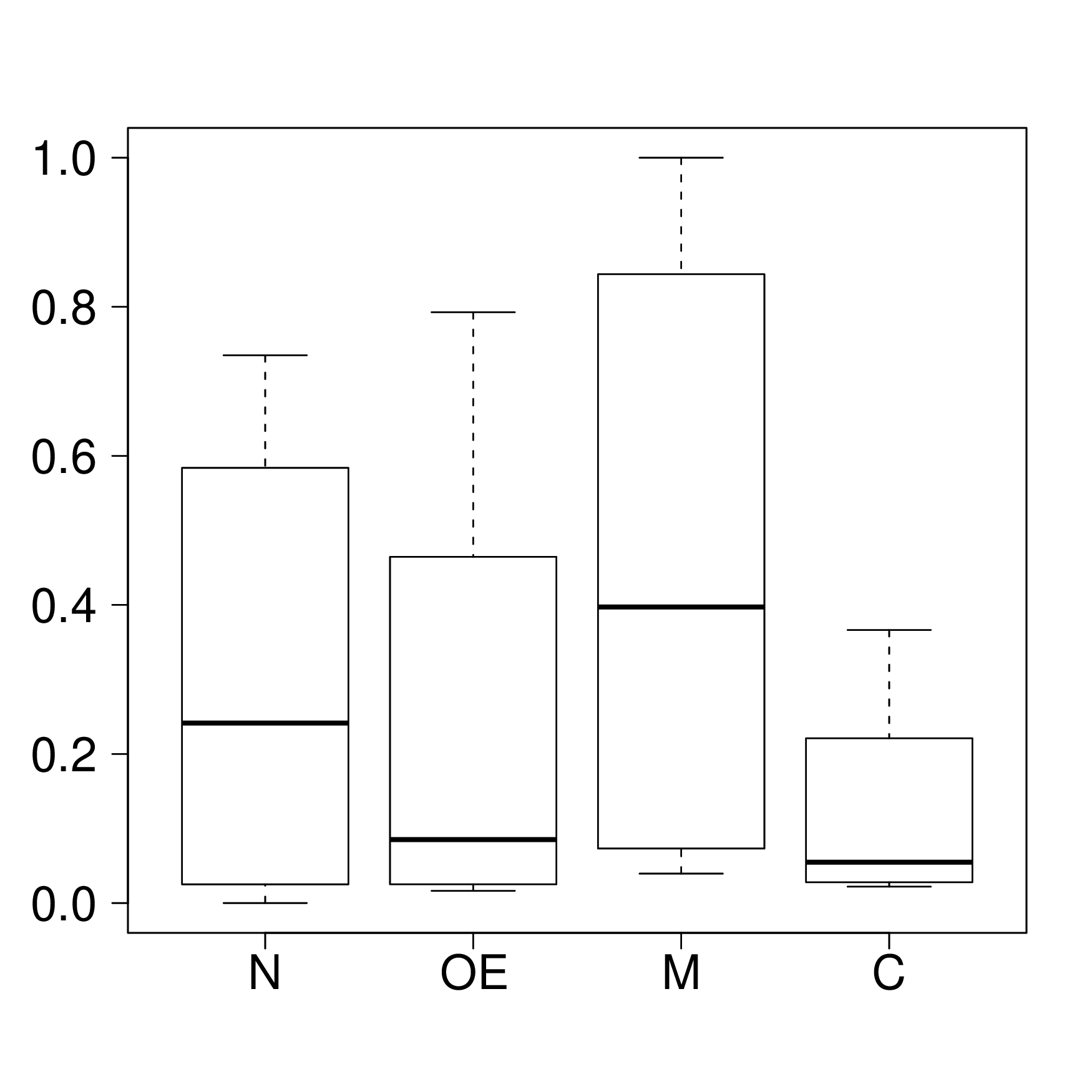}
  \label{fig:boxplot-links}
 }
 \subfigure[Topic repetition]{
  \includegraphics[width=90px]{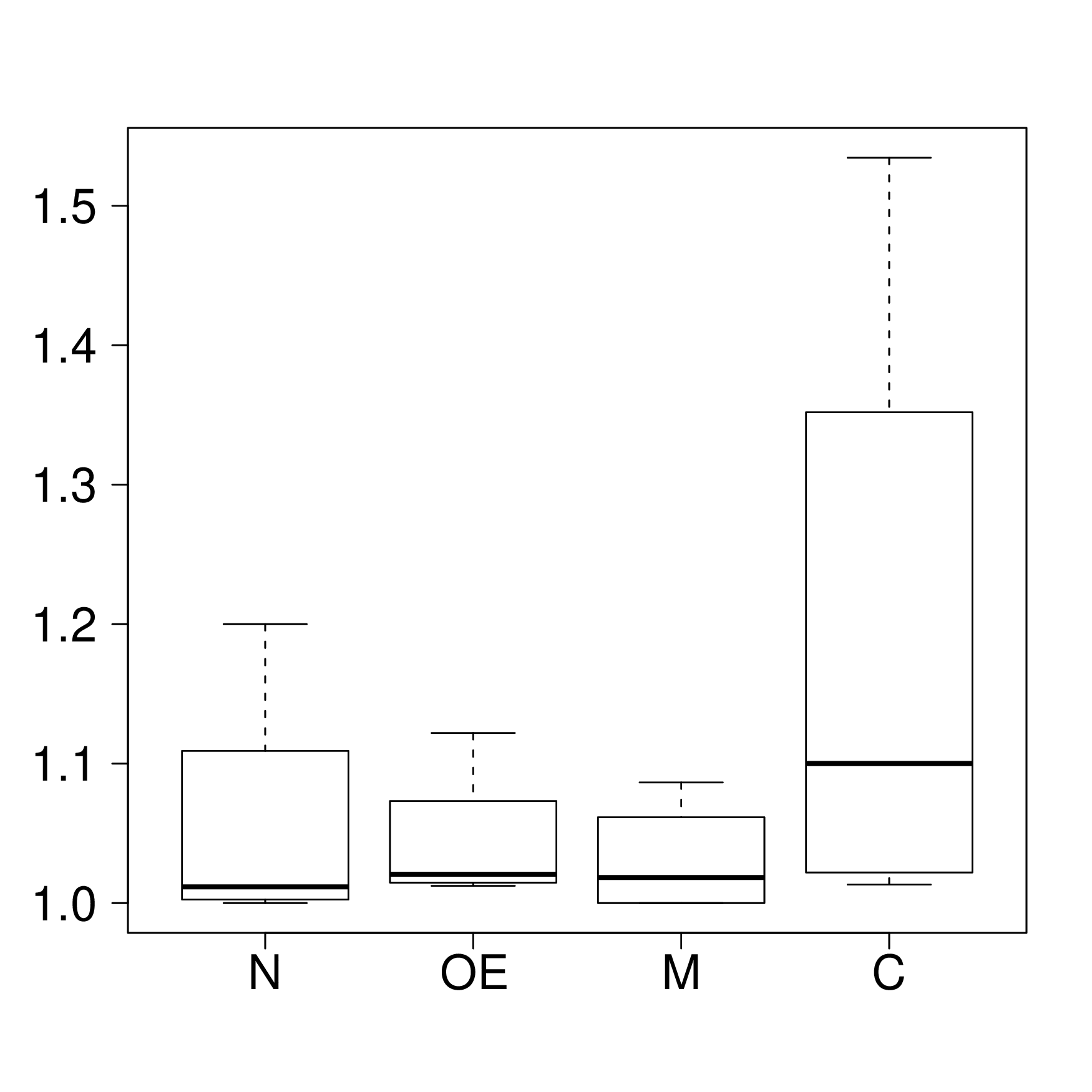}
  \label{fig:boxplot-query-repetition}
 }
 \subfigure[Replies]{
  \includegraphics[width=90px]{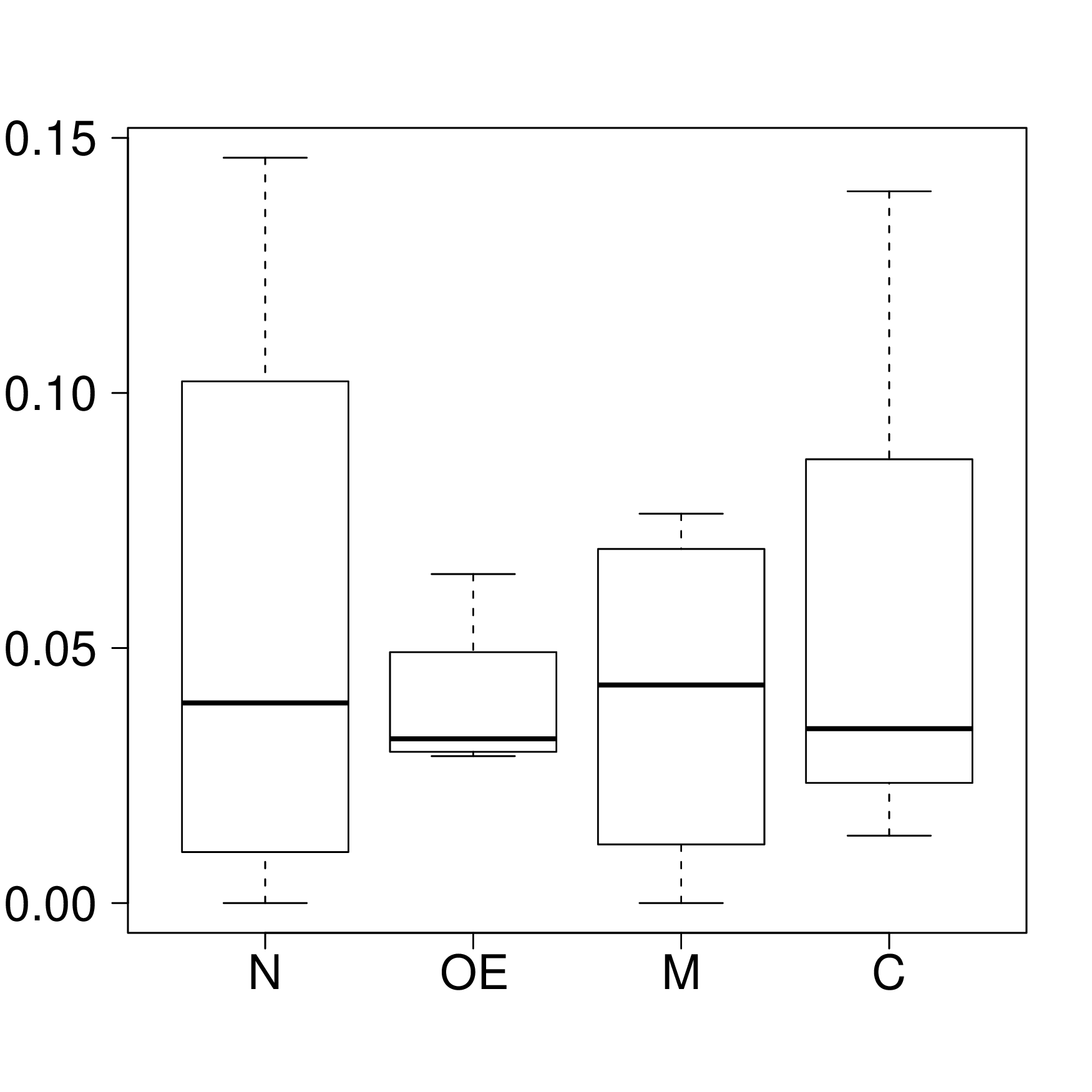}
  \label{fig:boxplot-replies}
 }
 \subfigure[Spread velocity]{
  \includegraphics[width=90px]{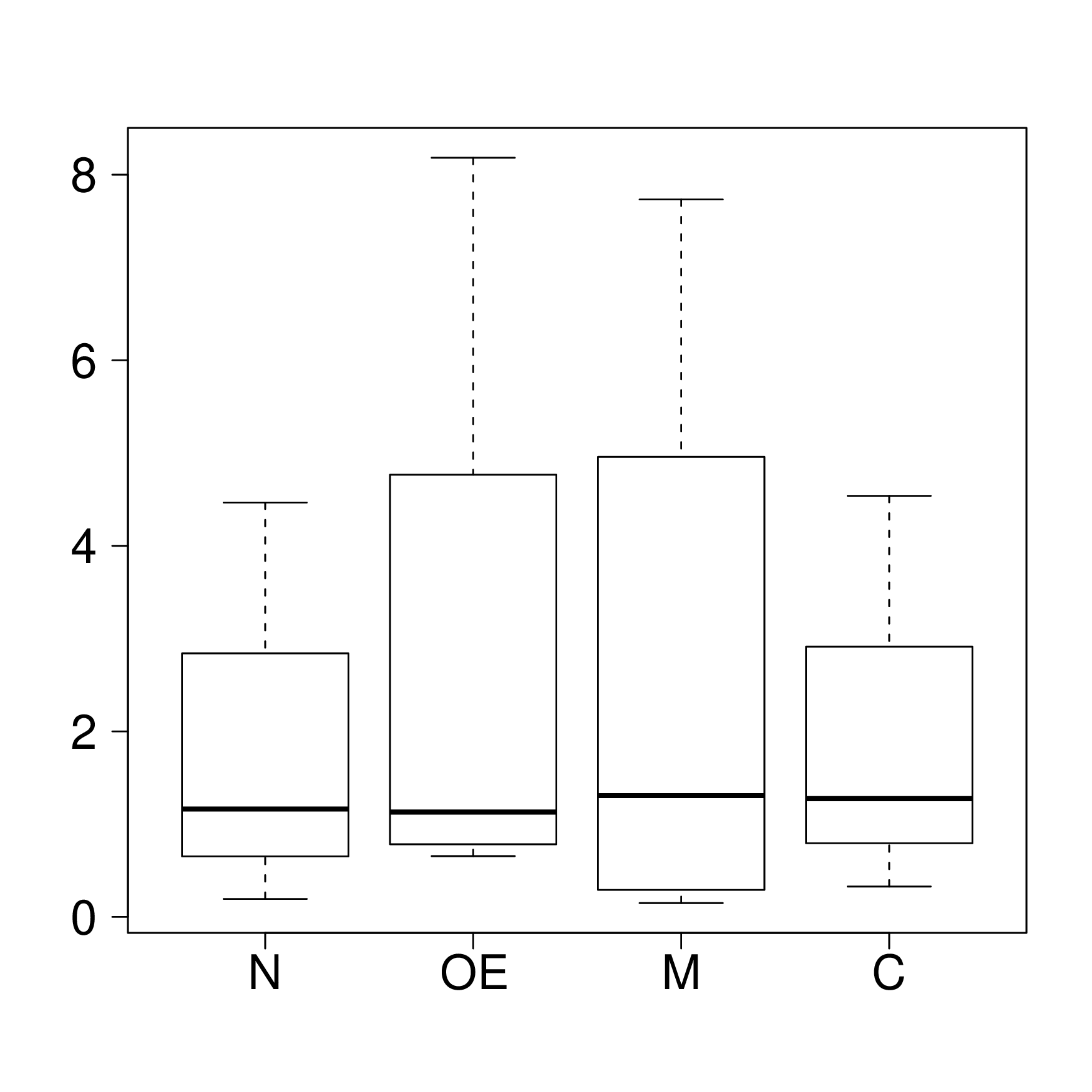}
  \label{fig:boxplot-spreadvelocity}
 }
 \caption{Box/whisker plots for features relying on average values (N: News; OE: Ongoing Events; M: Memes; C: 
Commemoratives).}
 \label{fig:average-feature-boxplots}
\end{figure*}

\begin{figure*}[htb]
 \centering
 \subfigure[User diversity]{
  \includegraphics[width=90px]{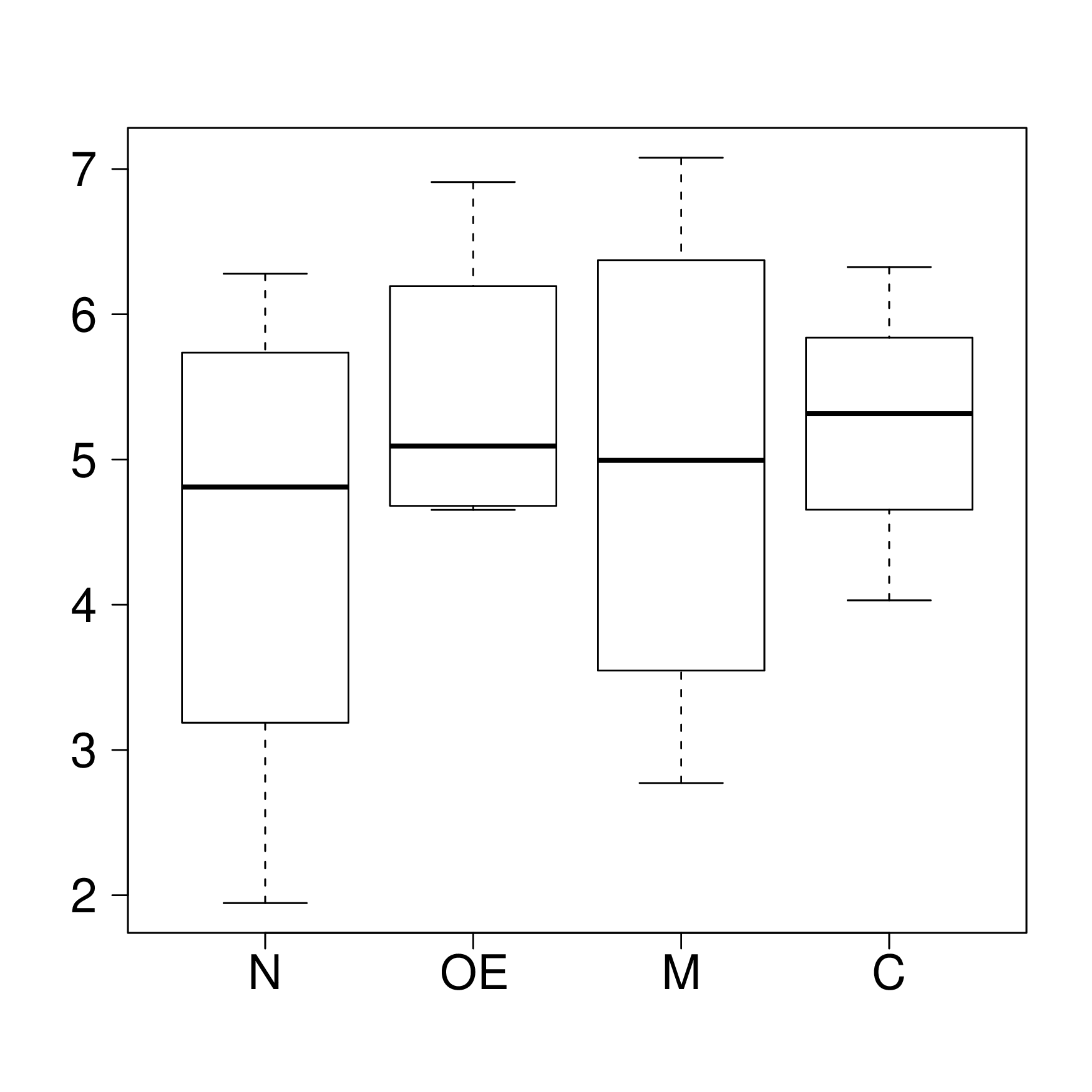}
  \label{fig:boxplot-userdiversity}
 }
 \subfigure[Retweeted user diversity]{
  \includegraphics[width=90px]{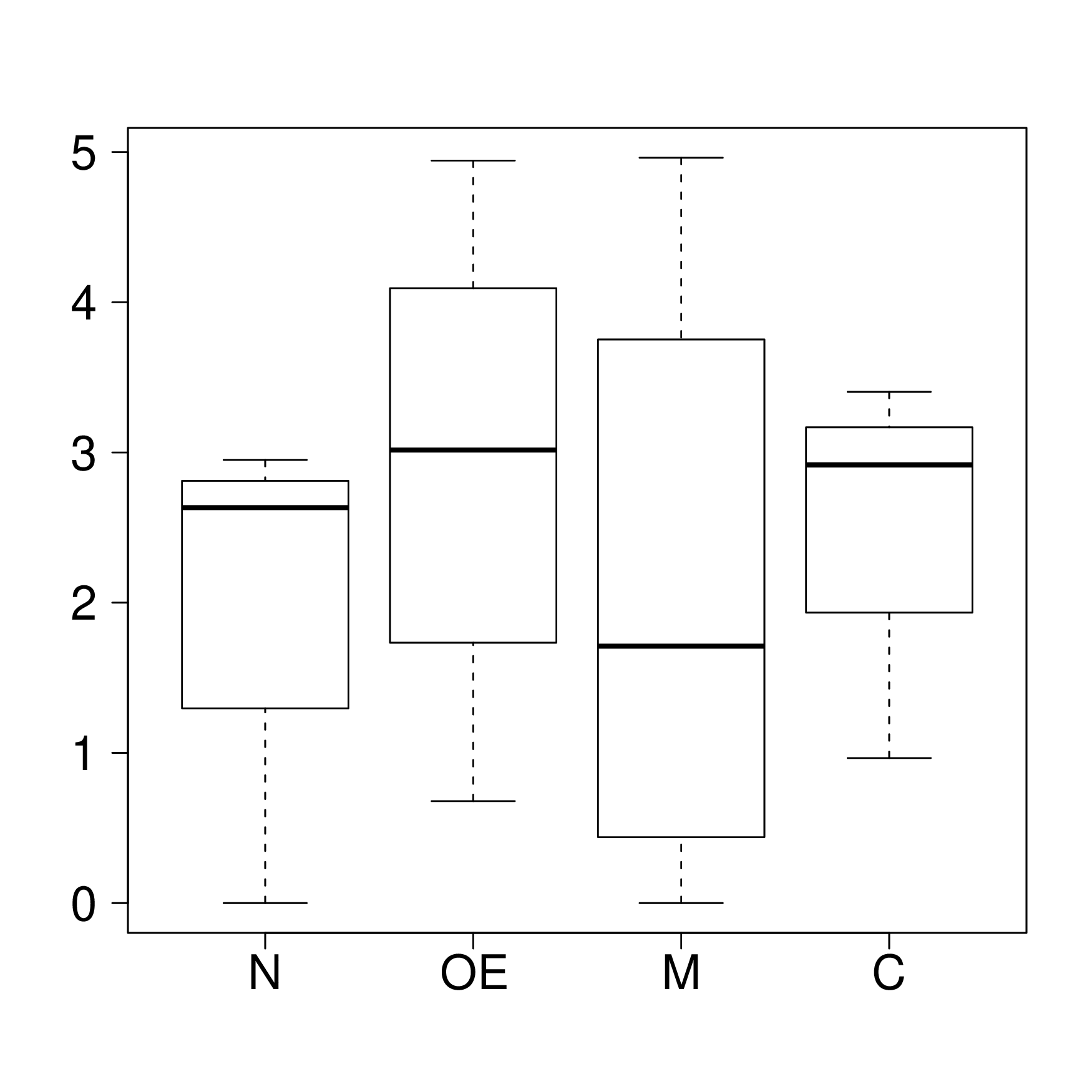}
  \label{fig:boxplot-retweeteduserdiversity}
 }
 \subfigure[Hashtag diversity]{
  \includegraphics[width=90px]{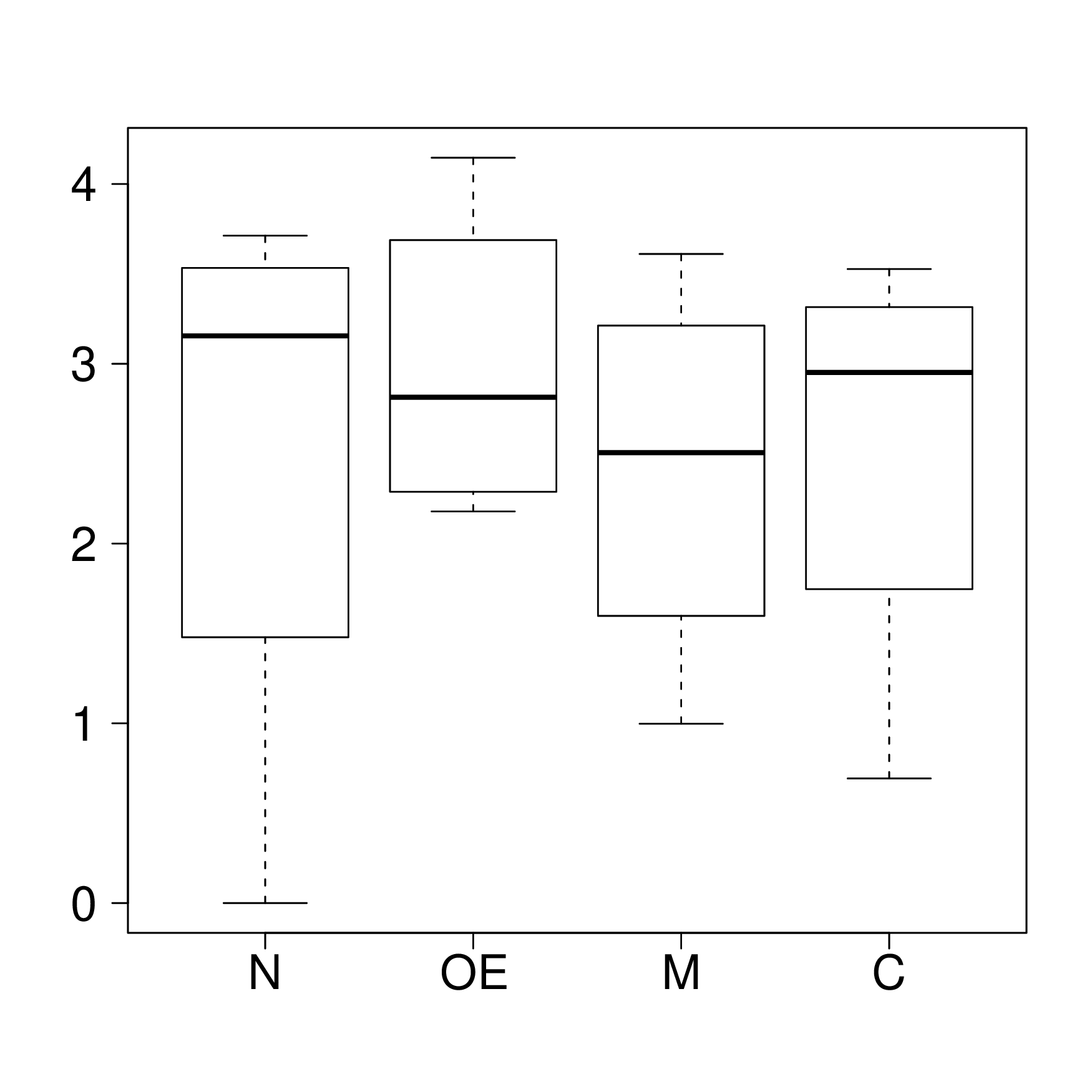}
  \label{fig:boxplot-hashtagdiversity}
 }
 \subfigure[Language diversity]{
  \includegraphics[width=90px]{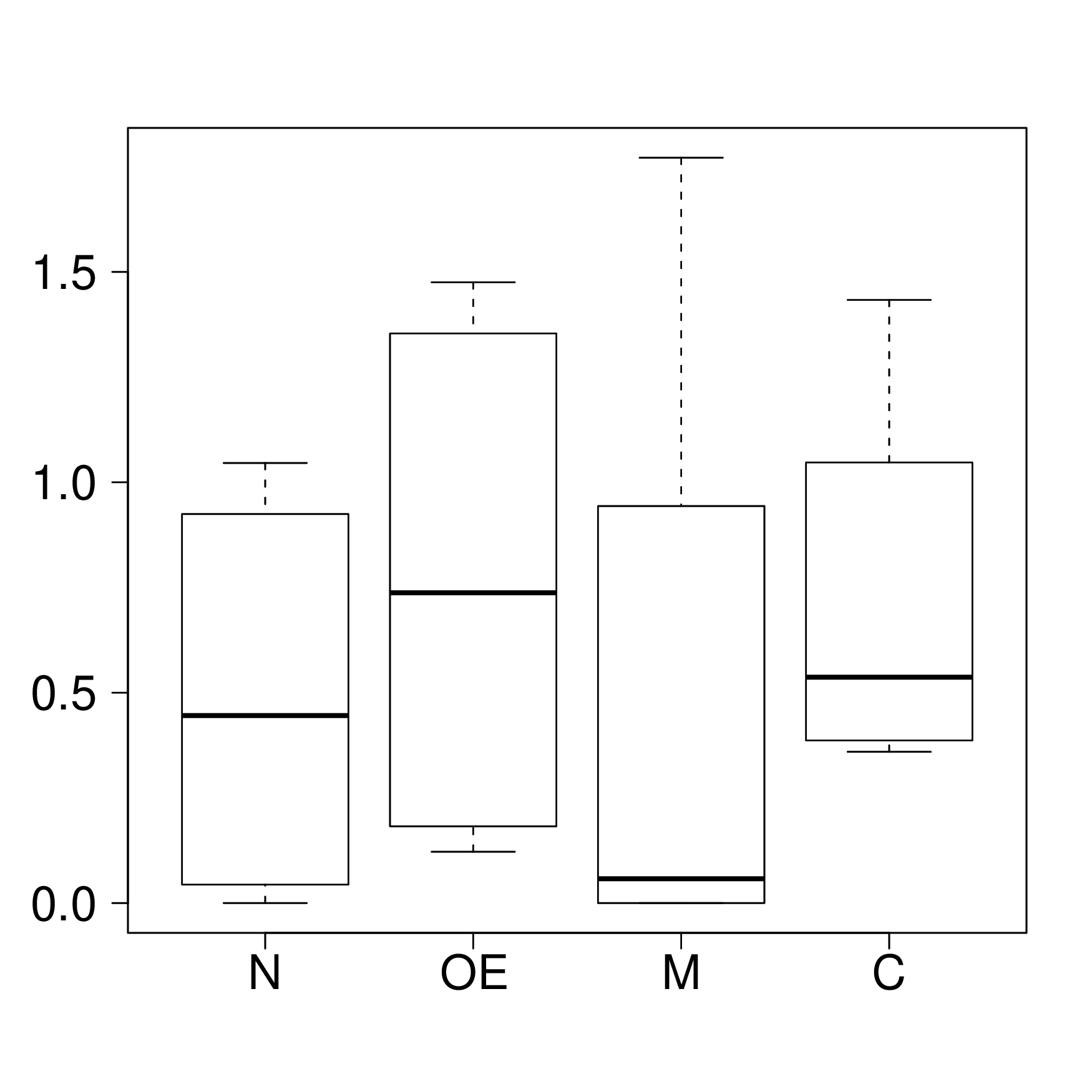}
  \label{fig:boxplot-languagediversity}
 }
 \subfigure[Vocabulary diversity]{
  \includegraphics[width=90px]{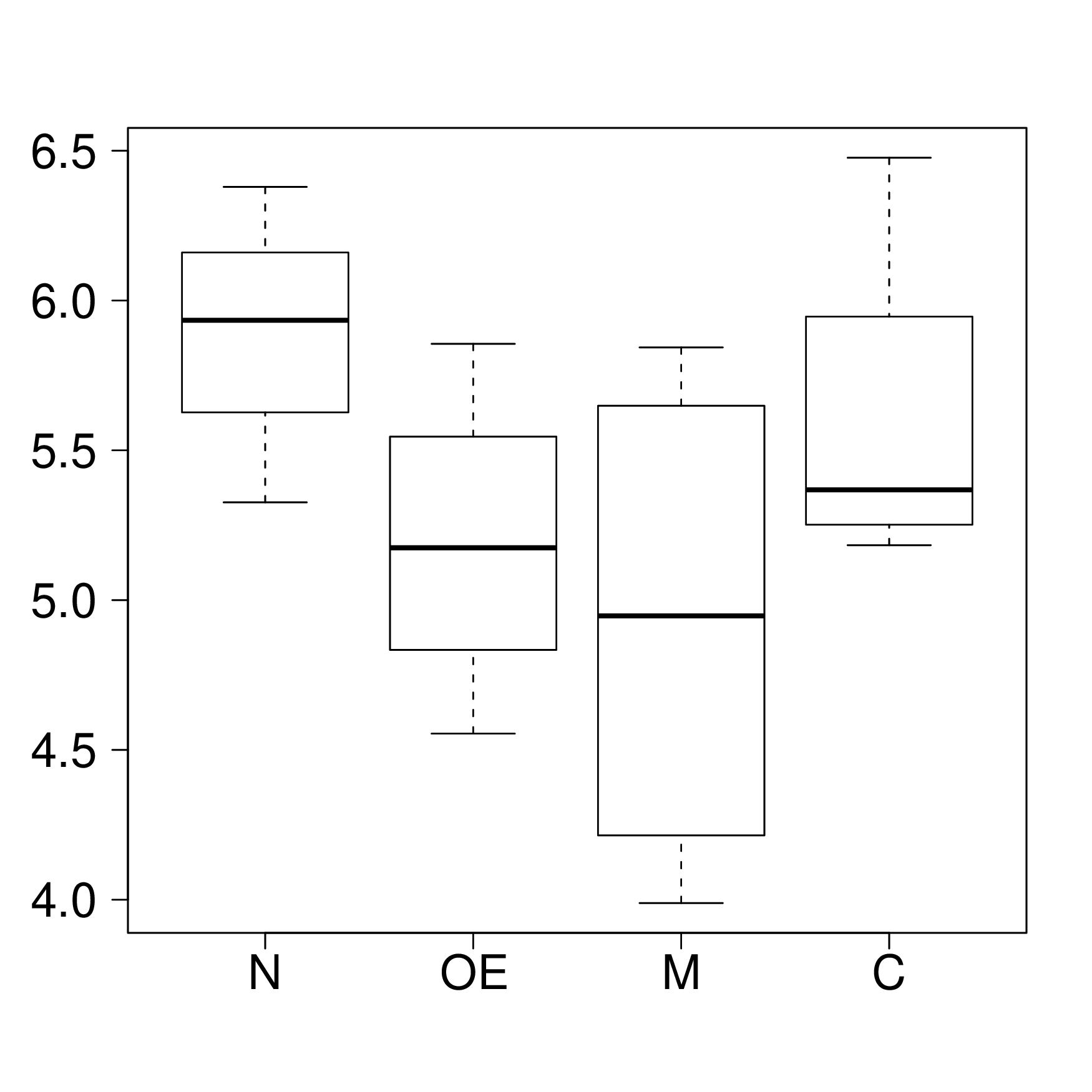}
  \label{fig:boxplot-vocabularydiversity}
 }
 \caption{Box/whisker plots for features relying on diversity values (N: News; OE: Ongoing Events; M: Memes; 
C: Commemoratives).}
 \label{fig:diversity-feature-boxplots}
\end{figure*}

Figures \ref{fig:average-feature-boxplots} and \ref{fig:diversity-feature-boxplots} show sets of box/whisker 
plots for average and diversity features, respectively. At a first glance, it can be seen that none of the 
features provide the same range of values for all the types of trending topics, so that all of them seem to 
be useful to some extent as to discriminating types. Analyzing them in more detail, some interesting thoughts 
can be inferred:

\begin{itemize}
 \item Figure \ref{fig:boxplot-retweets-ratio} and \ref{fig:boxplot-retweets} show that the use of retweets 
is more frequent for memes, and to a lesser extent for news. The use of retweets in memes seems to be due to 
the virality of this kind of trending topics.

 \item Figure \ref{fig:boxplot-retweets-ratio} showing the ratio of retweets contained in the trend shows 
that news and memes are retweeted most. Figure \ref{fig:boxplot-retweets}, which is very similar, shows that 
the retweeting depth of news and memes is very similar to the other, due to its high resemblance to the 
figure showing retweet ratios.

\item Figure \ref{fig:boxplot-hashtags} shows that, while the ratio of hashtags is very similar for 
breaking news and ongoing events, some of the memes and commemoratives have a higher ratio of hashtags. 
Interestingly, few commemoratives have tweets with more than one hashtag.

 \item Figure \ref{fig:boxplot-length} shows that tweets posted during ongoing events are significantly 
shorter than those in
the other classes. Intuitively, users who are tweeting about an ongoing event are likely to either be 
attending (e.g., a
conference) or watching (e.g., a TV show) an event. This suggests that while users are following another 
event, they devote less time to sending the tweet, either because of the lack of attention or time, or 
because they are using a hand-held device where writing requires more effort.

 \item Figure \ref{fig:boxplot-questions} shows that questions marks are used more frequently in 
commemorative trending topics. This suggests that users tweeting about commemorative issues are rather prone 
to asking about it. The use of questions marks is the lowest for news, where users are not so likely to ask, 
but rather get surprised because of the news (see the high use of exclamations in news in Figure 
\ref{fig:boxplot-exclamations}).

 \item Regarding the use of links in tweets, one could expect that news clearly had the highest density of 
links. However, Figure \ref{fig:boxplot-links} shows that memes have more links. This is due to the fact that 
several memes emerge from sharing links to funny content on the Web. On the other hand, the number of links 
in news is not that high. This corroborates the observation by \citet{kwak2010what} that sometimes news break 
earlier on Twitter than on the Web, thus not always there are links available as soon as the topic trends on 
Twitter. This strengthens the motivation to discover breaking news that trend on Twitter.

 \item Figure \ref{fig:boxplot-query-repetition} shows that the feature considering the repetition of the 
trending topic is not discriminative for news, ongoing events, and memes, but it presents a higher rate of 
repetition for commemorative trends.

 \item As regards to the spread velocity of the trends, Figure \ref{fig:boxplot-spreadvelocity} shows that 
there is some similarity among types, but some memes and ongoing events spread faster and have higher 
frequencies of tweets per second.

 \item Figure \ref{fig:boxplot-userdiversity} shows that the diversity of users contributing to the trending 
topic is uniformly distributed over the four classes. Nonetheless, it is worthwhile noting that the values 
are very low. This suggests that trending topics are usually generated by a small number of users. This 
corroborates the conclusion by \citet{asur2011trends} that a few users can be considered 
\emph{trend-setters}, who provide the seed that starts the trend, and loads of \emph{propagators} help spread 
it.

 \item Figure \ref{fig:boxplot-retweeteduserdiversity} shows the diversity of users who got re\-tweet\-ed in 
a trend. It can be seen that memes present the lowest diversity, followed by news, whereas the highest 
diversities occur in ongoing events and commemoratives. This means that the contribution of influential users 
is clearer for memes. Thus, there are fewer \emph{trend-setters} that spark the emergence of memes.

 \item Figure \ref{fig:boxplot-languagediversity} shows the diversity of languages of the tweets contained in 
trends. It shows that there are lots of memes with a low diversity of languages, where there is a predominant 
language in which most tweets were written. It seems that most tweets in memes are spread in the same 
language that it was started. The fact that memes are the most retweeted type of trend is relevant for this, 
since retweets retain the language of the original tweet. Ongoing events are the trends that show the highest 
diversity of languages, suggesting that many of the events are followed internationally --e.g., soccer games.

 \item Figure \ref{fig:boxplot-vocabularydiversity} shows higher vocabulary diversity for news and 
commemoratives. However, this higher diversity for news and commemoratives does not affect hashtags 
significantly, where diversity is comparable to the other types of trends (see 
Figure~\ref{fig:boxplot-hashtagdiversity}). This suggests that the increase of diversity is produced by 
content other than hashtags, which could be explained with other factors such as longer, wordier tweets for 
news.

\end{itemize}

\subsubsection*{Analysis of Text in Tweets}
\label{tweet-text-analysis}

In order to analyze the textual content of the tweets, we give an overview of the top terms occurring in each 
type of trending topic, i.e., the most frequent vocabulary used in each type of trending topic. To do so, we 
first performed a filtering process to remove irrelevant words. The filtering process removed all the 
stopwords contained in the tweets. The stopword removal process includes Twitter-specific words like 
\emph{RT}, and words in stopword lists for the main languages in the dataset. After that, we computed the TF 
(term frequency) of each word for each type of trending topic. This process produced a list of words for each 
type of trending topic, ranked in a descending order by TF value. Table \ref{tab:top-terms} shows a list of 
top 15 terms for each type of trending topic. It can be seen that several words in these lists have strong 
relation with the definitions of the types of trending topics:

\begin{itemize}
 \item For the commemorative trending topics, congratulating words stand out --e.g., \emph{happy} and 
\emph{birthday}--, as well as time-related words --e.g., \emph{day} and \emph{anos} (years in Portuguese)--, 
showing that most tweets have to do with celebrations and congratulations.
 \item For the trending topics related to ongoing events, several words refer to the present --e.g., 
\emph{tonight} and \emph{live}--, the action they are doing --e.g., \emph{watching} and \emph{watch}--, or 
the media they follow --e.g., \emph{tv}--. This shows that this kind of trending topic has some specific 
terminology associated, which is strongly related to happenings that are being live-tweeted.
 \item For the news-related trending topics, the terminology is not that specific. It seems that most of the 
top terms are named entities that appear frequently in the news. However, including the word \emph{news} 
itself seems frequent for this kind of trend, as it is the top term in this class.
 \item The terminology associated to memes seems to be the least specific. This makes sense because the range 
of topics covered by memes can be very wide, and it seems that the vocabulary of a meme can vary 
substantially. Note that a meme can be any of a funny thing, something that users are keen to happen or see, 
or an idea that unintentionally went viral.
\end{itemize}

Further, it seems that some terms tend to overlap across different classes. For instance, terms related to 
the actor Charlie Sheen appear as top terms for news, ongoing events, and memes. He had different trending 
topics in the time frame of our dataset: (1) users retweeted actor's message thanking the Twitter community 
for following him --meme--, (2) he broadcasted a live show on the Internet --current-event--, and (3) he was 
fired from the TV series \emph{Two \& a Half Men}.

\begin{table}[htbp]
 \begin{center}
  \begin{tabular}{  c  c  c  c }
   \toprule
   \textbf{N} & \textbf{OE} & \textbf{M} & \textbf{C} \\
   \midrule
   news & love & charliesheen & happy \\
   sheen & watching & winning & birthday \\
   charlie & lol & love & mamonas \\
   guttenberg & tonight & twitter & gainsbourg \\
   half & live & lol & assassinas \\
   men & good & day & bon \\
   twitter & video & lmaotwitpics & jon \\
   john & watch & justinbieber & jovi \\
   jane & david & video & day \\
   fired & game & devassa & anos \\
   russell & time & great & serge \\
   sky & tv & stone & bieber \\
   lina & idol & adriesubono & justin \\
   game & sheen & boy & jensen \\
   warner & jones & movie & ackles \\
   \bottomrule
  \end{tabular}
 \end{center}
 \caption{List of top terms in tweets grouped by type of trending topic (N: News; OE: Ongoing Events; M: 
Memes; C: Commemoratives).}
 \label{tab:top-terms}
\end{table}

\section*{Automatic Classification of Trending Topics}
\label{trend-classification}

Next, we describe the technical details of the classifier used in our experiments to classify trending 
topics. Afterward, we present the results of assessing the accuracy of the classifier, as well as the 
appropriateness of the typology introduced in this work.

\subsection*{Experimental Setup}
\label{experimental-setup}

In order to evaluate the suitability of classifying trending topics according to the established typology, we
perform classification experiments using Support Vector Machines (SVM) \citep{joachims98text}. Being a
multiclass task, we specifically rely on a \emph{one-against-all} binary combination method
\citep{rifkin2004defense}, implemented with 
\emph{svm-light}\footnote{\url{http://svmlight.joachims.org}}\footnote{We use the linear kernel and set 
the penalty parameter $C$ --which controls the trade off between allowing training errors and forcing rigid 
margins-- to 5, allowing a certain degree of misclassification. Those parameters were  found
suitable for another text classification task~\citep{zubiaga2009getting}, i.e., classification of web pages 
using metadata from social media.}. Instead of
considering the multiclass problem as a single task, \emph{one-against-all} splits it into smaller binary
ones. For a problem with \emph{k} classes, \emph{one-against-all} defines \emph{k} different classifiers. In
the training phase, each of the \emph{k} classifiers learns a model to separate a class from the rest
\textit{k-1}. This model creates a hyperplane to separate the class from the rest. In our task, where 4
classes are defined, the following classifiers are created: \textit{1 vs 2-3-4}, \textit{2 vs 1-3-4},
\textit{3 vs 1-2-4} and \textit{4 vs 1-2-3}. In the process of categorizing, each classifier provides an
output for each trending topic, which refers to the margin --i.e., distance to the hyperplane-- as a
reliability value. The classifier maximizing the output defines the final class predicted by the system (see
Equation \ref{eq:one-against-all-solution}).

\begin{equation}
 \hat{C}_{i} = \arg\max_{i = 1,...,k} m_i
 \label{eq:one-against-all-solution}
\end{equation}

where $m_i$ is the margin outputted by the classifier $i$.

We use two different representations in the classification process: the Twitter features analyzed 
above, and a bag-of-words approach of the textual content of tweets as a baseline. With both 
representations, we rely on the Vector Space Model (VSM), thus we define a vector for each trending topic. We 
create one vector per trending topic for each of the representations. We use a training set with 600 trending 
topics, with 436 in the test set. We perform 10 random selections from the dataset to form 10 training 
datasets so we show the average results in the tables. The two representations we use are the following:

\begin{itemize}
 \item For the Twitter features, we use vectors with 15 dimensions. The dimensions correspond to the social 
features introduced above. All the trends have a value for each of this dimensions, so that there are no 
features that remain without a value. The main advantages of this representation approach are the 
straightforwardness of the features, and that the number of features does not depend on the number of trends 
we have to represent, but it remains unchanged.

 \item For the textual content of tweets, we rely on the bag-of-words approach. We tokenize tweets and 
compute the number of occurrences of each term in a trend. Prior to that, we remove stopwords for the main 
languages in the dataset. Note that tweets have not been translated. Hence, the vocabulary used to represent 
the tweets includes the terms from all the languages in the collection. We represent each trend with the TF 
values of each term. This approach is computationally much more expensive than the approach based on Twitter 
features, and presents the problem that the number of dimensions utilized to represent the trends increases as 
the collection grows. In this case, representing the 1,036 trending topics generated vectors with 512,943 
dimensions.
\end{itemize}

Finally, we rely on the fact that the larger is the margin outputted by a classifier for a class, the more 
reliable can be considered the classifier's prediction on that class. Based on this idea, we use classifier 
committees \citep{sun_support_2004} to combine the outputs of classifiers based on Twitter features and the 
bag-of-words. Classifier committees add up the outputs of both classifiers. Note that, in this case, there 
are two classifiers (one based on Twitter features, and the other based on the textual content of tweets), 
and four classes as defined in the typology. Thus we can get the sum of the outputs from both classifiers for 
each class. Afterward, the class with the highest sum value is selected as the prediction of the committees 
(see Equation \ref{eq:committees-prediction}). 

\begin{equation}
 C^*_{t} = \arg \max_{i = 1..k} \{S_{ti}\}
 \label{eq:committees-prediction}
\end{equation}

where $k$ is the number of classes, $t$ is the trending topic, and $S_{ti}$ is the sum given by Equation 
\ref{eq:sum-margins}.

As an example of the possible advantage of using classifier committees, Table \ref{tab:example-committees} 
shows the outputs from the two classifiers for a trending topic. Even though one of the classifier labels the 
trending topic as news, and the other labels it as meme, adding up outputs using classifier committees 
changes the label to ongoing event. Classifier committees can be suitable in some cases to get the most from 
both classifiers.

\begin{table}[htb]
\begin{center}
 \begin{tabular}{|l|c|c|c|c|}
  \hline
  & \textbf{N} & \textbf{OE} & \textbf{M} & \textbf{C} \\
  \hline
  \hline
  \textbf{Classifier A} & \textbf{1.2} & 1.1 & 0.6 & 0.3 \\
  \hline
  \textbf{Classifier B} & 0.5 & 1.0 & \textbf{1.2} & 0.3 \\
  \hline
  \hline
  \textbf{Classifier committees} & 1.7 & \textbf{2.1} & 1.8 & 0.6 \\
  \hline
 \end{tabular}
\end{center}
\caption{Example of classifier committees on a trending topic. Classifier A labels it as news, and classifier 
B labels it as meme, while adding both up using classifier committees would label it as an ongoing event.}
\label{tab:example-committees}
\end{table}

\begin{equation}
 S_{ti} = \sum_{v=1}^{|V|} m_{vti}
 \label{eq:sum-margins}
\end{equation}

where $|V|$ is the number of classifiers, and $m_{vti}$ is the margin outputted by the classifier $v$ for the
trending topic $t$ in the class $i$.

\subsection*{Evaluation}
\label{evaluation}

\begin{table}[htbp]
 \begin{center}
  \begin{tabular}{  l  c  c }
   \toprule
   \multicolumn{1}{l}{}& \textbf{Accuracy} & \textbf{Kappa}\\
   \midrule
   \textbf{Bag-of-words} & .752 & .530 \\
   
   \textbf{Twitter features} & .784 & .604 \\
   \textbf{Committees} & .812 & .649\\
   \bottomrule
  \end{tabular}
 \end{center}
 \caption{Accuracy and Cohen's Kappa statistic of the trend classification by type of representation.}
 \label{tab:accuracy}
\end{table}

\begin{table}[htbp]
 \begin{center}
  \begin{tabular}{  l  c  c  c  c }
   \toprule
   \multicolumn{1}{l}{} & \multicolumn{4}{c}{\textbf{Class Precision}}  \\
   \multicolumn{1}{l}{}  & \textbf{N} & \textbf{OE} & \textbf{M} & \textbf{C} \\
   \midrule
   \textbf{Bag-of-words} & .673 & .783 & .636 & .548 \\
   \textbf{Twitter features} & .634 & .829 & .731 & .132 \\
   \textbf{Committees} & .598 & .941 & .687 & .200 \\
\bottomrule
  \end{tabular}
 \end{center}
 \caption{Precision by class (N: News; OE: Ongoing Events; M: Memes; C: Commemoratives) of the trend 
classification by type of representation.}
 \label{tab:precision}
\end{table}

Table \ref{tab:accuracy} shows the accuracy values of the classification of trending topics. The accuracy 
measures the percent of correct guesses among all the predictions. The representation based on the proposed 
Twitter features achieves a superior accuracy than the bag-of-words baseline. This superiority gap is of 
3.2\% in favor of the representation based on Twitter features. Furthermore, the winning approach presents 
the advantages that it only requires 15 features instead of the thousands required by the content-based 
representation and, the features are straightforward and easy to compute. When analyzing in more depth the 
precision by class (see Table \ref{tab:precision}), it can be seen that Twitter features perform better in 
the detection of ongoing events, and especially memes, where the gap is bigger than 9\%. Using the 
bag-of-words improves for news, and especially for commemoratives, where Twitter features get low precision. 
However, this could be due to the small number of commemoratives 
in the collection\footnote{Note that there are only 27 commemorative trends in the collection.}, and it may 
perform better with more trends of this type, probably providing a better representation of the class in the 
training stage.

The use of classifier committees combining the outputs of both classifiers yield even superior results. It 
improves the performance of the use of Twitter features by 2.8\%. As \citet{sriram2010short} report for the 
classification of single tweets, our work also shows that the combination of bag-of-words with Twitter 
features yields better results in the classification of trending topics. However, this outperformance comes 
especially because of the high performance of the committees on ongoing events. Single classifiers perform 
better than the committees for the rest of the classes. Moreover, it requires both classifiers --Twitter 
features and bag-of-words-- to be run first, in order to perform the combination afterward. The use of 
committees can be a good option to detect ongoing events when immediacy is not a key requirement.

Due to the fact that classes are unbalanced, accuracy values may be subject to guesses given by randomness. 
Thus, the accuracy does not always compensate for hits that can be attributed to randomness 
\citep{BenDavid2007875}. Because of this, we also show the Cohen's Kappa statistics~\citep{cohen1960coefficient} to measure 
the 
extent to which accuracy values are independent of randomness (see Equation \ref{eq:kappa}).

\begin{equation}
 \kappa = \frac{P_0 - P_c}{1 - P_c}
 \label{eq:kappa}
\end{equation}

where $P_0$ is the relative observed agreement between the predictions of a classifier and the observed 
categorization (accuracy), and $P_c$ is the hypothetical probability of chance agreement, using the observed 
data to calculate the probabilities of each observer randomly saying each category. If the categorizations 
are in complete agreement then $\kappa$ = 1. If there is no agreement between the categorizations, then 
$\kappa$ = 0.

The kappa values in Table \ref{tab:accuracy} show the impact of randomness on the agreement between predicted 
and observed categorizations for the 3 classifiers. Note that the higher is the value, the lower is the 
effect of randomness on the accuracy improvement. It can be observed that the kappa values follow the same 
rank as the accuracy values, so that it indicates that the accuracy improvement is not a consequence of 
random guesses.

\section*{Discussion}
\label{discussion}

This study attempts to satisfy the dearth of research analyzing, characterizing, and classifying social 
trends in their early stage, with the aim of feeding systems that can benefit from real-time knowledge. As 
such, it is not only relevant for researchers studying trends in social media, but also for developers who 
are eager to enhance their systems with real-time data from social networks. The typology of trends and 
methodology introduced in this paper enable us to serve the social trends of interest to different audiences: 
(i) the identification of trends categorized as \textit{news} can help discover breaking news early on, which 
would considerably facilitate the task of tracking social media for journalism practitioners, which has 
become paramount in their daily newsgathering process \citep{zubiaga2013curating}, (ii) the identification of 
\textit{ongoing events} can feed an event summarization system (e.g., \citet{zubiaga2012towards}), which 
would quickly be informed that a new event of wide interest is 
going on and is potentially worth being summarized, (iii) the identification of \textit{memes} can be a 
powerful input to marketing decisions \citep{williams2000business}, to find out the topics of interest and 
viral concepts that succeed at the very moment, as well as for researchers in the social sciences, to study 
the virality of ideas online, and (iv) the identification of \textit{commemorative} trends, which might often 
be known beforehand, might be interesting to mine and analyze the way they emerge, how they are spread, and 
what opinions or comments users associate with them \citep{pak2010twitter}. Here we have listed a set of 
systems that can benefit from the output of a set of categorized social trends, which is not intended to be 
comprehensive, and we believe that it can be extended to numerous systems that can benefit from knowledge 
about the current status of the society.

While the typology of trends introduced in this work can be too cross-grained for certain applications, it 
can be easily extended by breaking down the categories into smaller subcategories. As an example from 
existing research, \citet{recuero2012fandoms} study a specific type of memes, which they call fandom memes. 
Fandom memes can be defined as the memes that are introduced by a large community of a celebrity's or groups 
fans. These specific memes can be categorized as a subcategory of memes in our typology, which could in turn 
be classified into more specific fandom memes depending on the type of fan, i.e., supporters of a political 
party, a soccer team, or a music band. Likewise, breaking news can be classified in terms of the type of 
diffusion, e.g., multiple contributors --as in the Arab spring-- or few authorative sources --CNN--. Rather 
than presenting an exhaustive low-level categorization of trending topics, this study pretends to be 
comprehensive in the analysis of features that characterize trends, 
to understand and be able to categorize them.

In this work, we relied on trending topics as provided by Twitter itself, controlling for the identification 
of trends on our end. Given that Twitter's algorithm for trend detection is not public, we cannot certify 
that this is the optimal solution. The use of alternative algorithms for trend detection would help quantify 
the effect of such step, which is not within the scope of this work. Besides that, as a future work we are 
planning to study how each type of trending topic evolves in the subsequent minutes/hours, especially to 
explore whether some trends evolve into a different kind of trend, e.g., how a widely known phenomenon such 
as the \textit{Harlem Shake}, which began as a meme, made it later to most news outlets in the form of a news 
story.

\section*{Related Work}
\label{related-work}

Twitter has become an essential platform for researchers studying social information sharing. Researchers 
have studied the microblogging phenomenon itself \citep{java2007why,kwak2010what,boyd2010tweet}, the 
information diffusion on the social network \citep{wu2011who,krishnamurthy2008few,yang2010modeling} as well as 
the content of tweets 
\citep{jansen2009twitter,sriram2010short,sakaki2010earthquake,hurlock2011searching,cheng2010you}.

Regarding classification of single tweets, \citet{sriram2010short} define a typology of five 
generic classes of tweets (news, events, opinions, deals, and private messages) in order to improve 
information filtering. The authors represent tweets using a small set of language-dependent features to 
classify tweets written in English. The use of these features outperforms the bag-of-words approach in the 
classification of tweets according to the typology.

\citet{castillo2011information} define a set of features associated to tweets, users, topics and the 
propagation of retweets in order to automatically classify news events on Twitter as credible or not 
credible, so as to detect misinformation or false rumors. They show the effectiveness of user and 
sentiment-based features to this end.

Little work has been done analyzing the properties of trending topics on Twitter. Most of them focus on event 
and topic detection \citep{asur2011trends,cataldi2010emerging,sakaki2010earthquake,albakour2013oair, 
zhao2012identifying} or summarization of scheduled events in real-time as the events unfold 
\citep{zubiaga2012towards}. \citet{asur2011trends} explore the longevity of trending topics on 
Twitter, and analyze 
the role of users in the emergence of trends. They found that (i) a few users are \emph{trend-setters}, i.e, 
early contributors in the emergence of a trend, and (ii) a larger group of users are \emph{propagators}, who 
help spread the topic. \citet{cheong2009integrating} and \citep{kwak2010what} analyze 
the evolution of trending topics over the time, and perform a qualitative study of social features that 
characterize trending topics.

\citet{cui2012discover} categorize hashtags to discover vocabulary associated with news, filtering out 
memes, idioms and advertisements. They make use of features of hashtags based on temporal behavior, authors' 
distribution and word splitting to capture meme probability over popular hashtags in a six-month Twitter 
dataset.

\citet{recuero2012fandoms} study the strategies used by fans and anti-fans of music bands to 
generate trending topics. This type of trending topics can be considered a particular type of memes. They 
define five types 
of fan-based trending topics: Tribute, Promotion, Response, Request and War.

 \citet{kairam2013towards} study how to engage users of a search engine with social media when querying 
about trending events. They achieve a substantial agreement categorizing a sample of 99 
trending events --from Twitter's trending topics and Bing's Popular Now search queries-- and considering the 
categories Breaking, Ongoing, Meme, Commemorative and Unknown.

Complementary to our trending topics' taxonomy, \citet{naaman2011jasist} define a taxonomy of 
trending topics, focusing on characterizing local events (e.g., in a city) as endogenous trending topics -- 
those that emerge within Twitter and do not correspond to external events -- from exogenous trending topics 
-- 
that is, trends caused by real events that propagate through social media. As the authors explain, a trend is 
represented with a long time window (tweets sharing the trending topic terms 72 hours before and after the 
trend's peak, 1k tweets in average).

Different from the above, our work considers the following novel aspects that, as far as we know, had not been 
addressed before: (i) we define a high level typology that encompasses Twitter's worldwide trending topics, 
and (ii) we describe a system and features that accurately identify, in real-time, the type of trigger that 
sparks a trending topic, instead of considering the whole lifetime of a trend.

\section*{Conclusions}
\label{conclusions}

In this work, we have explored the types of triggers that spark conversations on Twitter. Specifically, we 
have explored the top conversations shown as trending topics on the site. We have introduced a typology to 
organize Twitter's trending topics by the type of happening that caused them. This typology includes the 
following 4 types of trending topics: \emph{news}, \emph{ongoing events}, \emph{memes}, and 
\emph{commemoratives}. We aimed at characterizing trending topics so that we were able to organize them by 
type. To this end, we have set out 15 straightforward features to represent trending topics. These features 
are independent of the language used in the tweets, and rely on the social diffusion of the trending topic. 
The features are of two different types: averages, and diversity values.

We have performed classification experiments using SVM classifiers to study the usefulness of these features 
to discriminate types of trending topics. We have shown that the proposed features provide more accurate 
classification results than the use of textual content of tweets. Furthermore, these features can 
straightforwardly be computed, and only require tweets sent before the topic trended, without need of external 
data. The proposed method provides an immediate way to accurately organize trending topics using a small 
amount of features. Unlike for the content-based representation, the number of features remains unchanged, and 
does not increase as the collection grows.

We have also analyzed and studied the nature and characteristics of the proposed features. The qualitative 
analysis of features provides deeper insight to understand how different happenings spread on social media. We 
have proven that the proposed features can help discriminate the type of trigger that caused the trending 
topic.

\section*{Acknowledgments}

This work has been part-funded by the Spanish Ministry of Education (FPU grant nr AP2009-0507), the
Spanish Ministry of Science and Innovation (Holopedia Project,
TIN2010-21128-C02), the Regional Government of Madrid and the ESF
under MA2VICMR (S2009/TIC-1542) and the European Community's FP7 Programme
under grant agreement nr 288024 (LiMoSINe).

\bibliographystyle{kbib_custom}
\bibliography{tt-jasist2013}

\end{document}